\documentclass[pra,twocolumn,superscriptaddress]{revtex4}

\usepackage[utf8x]{inputenc}
\usepackage[pdftex]{graphicx}
\usepackage{amsmath,amssymb}
\usepackage{subfigure}
\usepackage{todonotes}
\usepackage{bbold}

\newlength{\graphwid} 
\def\showgraph#1#2{
\settowidth{\graphwid}{\includegraphics[#1,clip=true]{#2}}
\parbox[c]{\graphwid}{\includegraphics[#1,clip=true]{#2}}}

\begin{document}

\title{Green's Function Formalism for Waveguide QED Applications}

\author{Michael P. Schneider} 
	\thanks{These two authors contributed equally}
	\affiliation{Max-Born-Institut, Max-Born-Str. 2A, 12489 Berlin, Germany}

\author{Tobias Sproll}
 	\thanks{These two authors contributed equally}
	\affiliation{Max-Born-Institut, Max-Born-Str. 2A, 12489 Berlin, Germany}

\author{Christina Stawiarski} 
	\affiliation{Institut f\"ur Theorie der Kondensierten Materie and 
	             DFG-Center for Functional Nanostructures,
               Karlsruher Institut f\"ur Technologie, 76128 Karlsruhe, Germany}

\author{Peter Schmitteckert} 
	\affiliation{Institut f\"ur Theorie der Kondensierten Materie and 
	             DFG-Center for Functional Nanostructures,
               Karlsruher Institut f\"ur Technologie, 76128 Karlsruhe, Germany}
  \affiliation{Institut f\"ur Nanotechnologie, 
	             Karlsruher Institut f\"ur Technologie, 76021 Karlsruhe, Germany}

\author{Kurt Busch} 
	\affiliation{Max-Born-Institut, Max-Born-Str. 2A, 12489 Berlin, Germany} 
	\affiliation{Humboldt-Universit\"at zu Berlin, Institut f\"ur Physik, 
	             AG Theoretische Optik \& Photonik, Newtonstr. 15, 12489 Berlin}

\begin{abstract}
We present a quantum-field-theoretical framework based on path integrals and Feynman 
diagrams for the investigation of the quantum-optical properties of one-dimensional 
waveguiding structures with embedded quantum impurities. In particular, we obtain the 
Green's functions for a waveguide with an embedded two-level system in the single- and 
two-excitation sector for arbitrary dispersion relations both in the time and the 
frequency domain. In the single excitation sector, we show how to sum the diagrammatic 
perturbation series to all orders and thus obtain explicit expressions for physical 
quantities such as the spectral density and the scattering matrix. In the two-excitation 
sector, we show that strictly linear dispersion relations exhibit the special property
that the corresponding diagrammatic perturbation series terminates after two terms, again
allowing for closed-form expressions for physical quantities. In the case of general 
dispersion relations, notably those exhibiting a band edge or waveguide cut-off frequencies, 
the perturbation series cannot be summed explicitly. Instead, we derive a self-consistent 
$T$-matrix equation that reduces the computational effort to that of a single-excitation 
computation. This analysis allows us to identify a Fano resonance between the occupied
quantum impurity and a free photon in the waveguide as a unique signature of the few-photon
nonlinearity inherent in such systems. In addition, our diagrammatic approach allows for
the classification of different physical processes such as the creation of photon-photon
correlations and interaction-induced radiation trapping -- the latter being absent for
strictly linear dispersion relations. 
Our framework can serve as the basis for further studies that involve more complex 
scenarios such as several and many-level quantum impurities, networks of coupled waveguides, 
disordered systems, and non-equilibrium effects. 
\end{abstract}

\maketitle

\section{Introduction}
\label{sec:introduction}
Presently, research on nano-scale quantum-optical (NQO) systems is witnessing an increasing 
amount of attention worldwide and several distinct experimental approaches have been 
developed to a point where the realization of complex functional elements becomes 
feasible. Besides ordinary integrated optical waveguides and Photonic Crystals with
embedded quantum impurities 
\cite{OBrien-2009,Benson-2011,Benson-Wegener-2013}
these systems also include superconducting waveguide-QED settings
\cite{Nori-2011,Devoret-2013} 
and fiber systems with nearby trapped atoms
\cite{Mitsch-2014}. 
In view of the fact that efficient integrated single-photon sources 
\cite{Shields-2007,Claudon-2010,Laucht-2012, Fattahpoor-2013}
and advanced integrated single-photon detectors
\cite{Hadfield-2009,Pernice-2012,Sahin-2013,Najafi-2015}
are available, the design and control of few-photon nonlinearities moves more and more 
into the focus of research efforts
\cite{Chang-2014,Volz-2014}.
Such integrated sources, integrated detectors, and controlled few-photon nonlinearities
represent the basic building blocks for future integrated quantum information processing
technologies
\cite{Lodahl-2015}.

>From a theoretical angle, the transport of few photons in waveguiding systems and their
interaction with quantum impurities exhibits certain similarities with nano-electronic
transport problems so that several methodologies of electron transport theory have been
adapted to the analysis of NQO systems. These include (i) a Bethe-Ansatz approach 
\cite{ShenFan-2007-1}, 
the Lehmann-Symanzik-Zimmermann reduction technique
\cite{ShiSun-2009} 
and the Input-Output formalism
\cite{Fan10}
for determining the multi-particle scattering matrix and (ii) a Green's function approach 
has been developed that exploits the chirality of effective low-energy field theories which 
are derived from  the basic Hamiltonian (see Eqs. (\ref{eq:model:hamiltonian_realspace}) 
and (\ref{eq:model:hamiltonian_momentumspace}) below)
\cite{Pletyukhov-2012}.
These approaches have also been extended to the case of many photons
\cite{Zheng10,Zheng11} 
and more complicated quantum impurities
\cite{Zheng12}.
Furthermore, direct numerical approaches have been developed
\cite{Longo-2009, LongoSchmitteckertBusch-2010, Nysteen-2015}
some of them going beyond the rotating wave approximation that is implied with the 
aforementioned basic Hamiltonian
\cite{Burillo-2014}.

In the present work, we develop a versatile quantum-field-theoretical framework for the 
analysis of NQO systems that consist of combinations of photonic waveguiding elements
with embedded quantum impurities. Our framework is based on a path-integral formulation 
and the construction of associated Feynman diagrams. This facilitates the formulation of 
an efficient Green's function technique. We demonstrate the efficiency of our framework
by rederiving the main results of the above-mentioned approaches and extending them to
arbitrary dispersion relations, notably for the case of two photons. In addition, our
approach provides novel and detailed insights into the physical processes that underlie
photon correlations effects and the existence of bound photon-atom states in such
systems. 

The manuscript is organized as follows. In Sec.~\ref{sec:model}, we describe the basic 
physical model and provide the corresponding path-integral formulation in Sec.~\ref{sec:pathint}.
This leads to a representation in Feynman diagrams which we elaborate on in Sec.~\ref{sec:FeynmanDiagrams}.
As Green's function are the central element of any diagrammatic approach, we discuss their
properties in the single- and two-excitation sectors in Secs. \ref{sec:spcase} and 
\ref{sec:2pcase}, respectively. The results are summarized in Sec.~\ref{sec:Conclusion} and
several technical aspects are relegated to appendices.

\section{The model}
\label{sec:model}

\begin{figure}[tb]
 \begin{center}
  \includegraphics[width=0.45\textwidth]{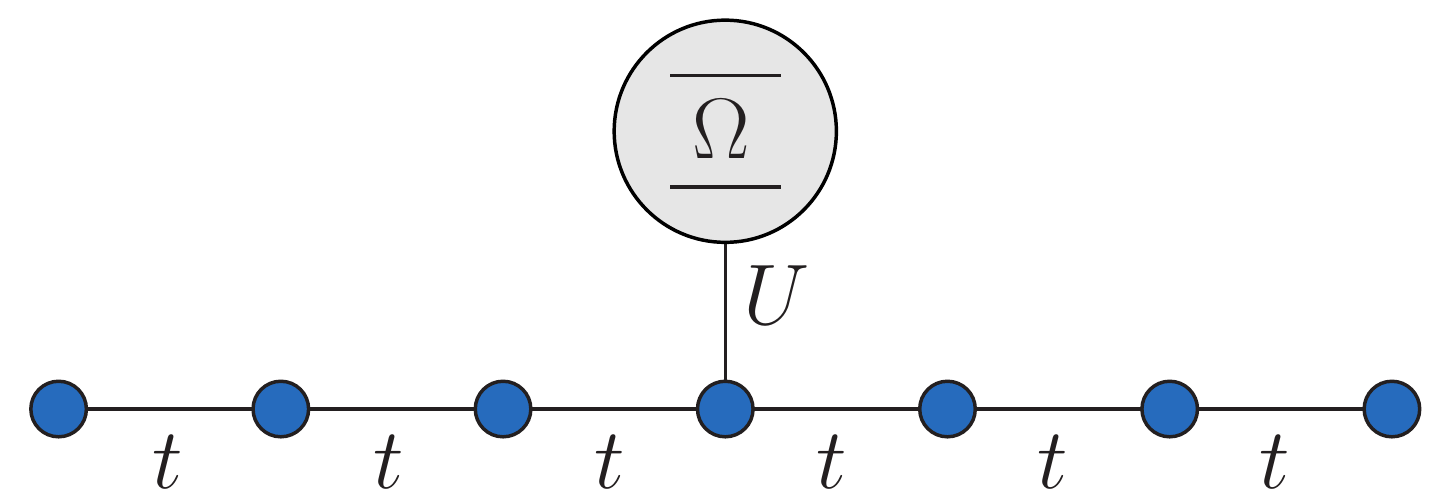}
 \end{center}
 \caption{Graphic representation of the model considered in this work: A tight-binding 
          bosonic chain with interchain hopping $t$ (representing a photonic waveguide)
          is site-coupled to a fermionic two-level system with level spacing $\Omega$
          and the few-photon transport through this system is studied.}
 \label{fig:model:model}
\end{figure}
We consider a one-dimensional bosonic quantum wire formed by a chain of sites to which a 
(fermionic) two-level system (TLS) is side-coupled, as sketched in Fig.~\ref{fig:model:model}. 
The corresponding Hamiltonian is given by

\begin{equation}\label{eq:model:hamiltonian_realspace}
	H = - t \sum_x \left( a^{\dagger}_{x} a^{\phantom{\dagger}}_{x+1} + \text{h.c.} \right) 
	    + \frac{\Omega}{2} \sigma_z + U \left( a^{\dagger}_0 \sigma_- + \text{h.c.} \right),
\end{equation}
where $a^{\dagger}_x$ is a bosonic creation operator at site $x$, $t$ is the in-chain hopping 
constant between nearest-neighbor sites of the wire. The TLS is described by the level spacing 
$\Omega$ and the corresponding Pauli matrix $\sigma_z$. The coupling (with strength $U$) between 
the quantum wire and the TLS is facilitated by the product of the TLS lowering operator $\sigma_-$ 
and the photon creation operator $a^{\dagger}_x$. Further, $\text{h.c.}$ denotes hermitian conjugation
and all lengths have been scaled by the lattice spacing $a$. This prototypical Hamiltonian
thus describes the propagation of photons in a waveguide that interact with an embedded quantum
impurity, the TLS, within the rotating wave approximation. At this point, we would like to 
note that the above bosonic sites could be regarded as identical physical resonators that 
are coupled through overlapping field distributions. Such systems have been fabricated
in coupled-resonator 
\cite{Notomi-2008},
photonic crystal- \cite{Notomi-2013} 
and fiber-optical 
\cite{Schell-2015}
settings
and their cosine-type dispersion relation has indeed been observed. Alternatively (and more
generally), these bosonic sites could be regarded as the numerical discretization of an 
arbitrary waveguide with a given dispersion relation. This dispersion relation can be
modeled by going beyond the nearest-neighbor hopping in the bosonic chain. However, typical 
waveguide dispersion relations usually feature (i) slow-light regimes in the vicinity of 
band edges or waveguide cut-off frequencies and/or (ii) frequency ranges with an almost linear 
dispersion relation. 
The tight-binding model, too, exhibits these features for frequencies near the edges and in 
the middle of the band, respectively, so that this model may serve as a good approximation 
for specific frequency ranges of general dispersion relations. 

Upon spatially Fourier-transforming Eq.~\eqref{eq:model:hamiltonian_realspace}, we obtain
\begin{equation}\label{eq:model:hamiltonian_momentumspace}
	H = \sum_k \epsilon(k) a^{\dagger}_{k} a^{\phantom{\dagger}}_{k} 
	    + \frac{\Omega}{2} \sigma_z 
	    + \frac{U}{\sqrt{L}} \sum_{k} \left( a^{\dagger}_k \sigma_-  
	    + \text{h.c.} \right),
\end{equation}
where $L$ is the system size and $\epsilon(k) = -2 t \cos(k)$ is the dispersion relation of 
the tight-binding chain. Note that the above Hamiltonian conserves the excitation number
\begin{equation}\label{eq:model:excitationnumber}
 N = \sum_{k} a_{k}^{\dagger} a^{\phantom{\dagger}}_{k} + \frac{1}{2} \left( \sigma_z + 1 \right),
\end{equation}
which means that the Hilbert space factorizes into subspaces of constant excitation number 
(as a matter of fact, we will only consider the cases $N=1$ and $N=2$ here). As a last step, 
we replace the Pauli spin operators by auxiliary fermions,
\begin{align}
 \sigma_z &= f^{\dagger} f - g^{\dagger} g \\
 \sigma_- &= g^{\dagger} f,
\end{align}
which requires that the constraint 
\begin{equation}
 f^{\dagger}f + g^{\dagger}g = 1,
\end{equation}
has to be fulfilled.
Here, $f^{\dagger}$ and $g^{\dagger}$ are, respectively, the creation operators of the excited 
and ground state of the TLS. Hence, the Hamiltonian which we use for the remainder of this 
work is 
\begin{align}\label{eq:model:hamiltonian}
	H = & \sum_k \epsilon(k) a^{\dagger}_{k} a^{\phantom{\dagger}}_{k} 
	      + \frac{\Omega}{2} \left( f^{\dagger} f - g^{\dagger} g \right) \nonumber \\
      & + \frac{U}{\sqrt{L}} \sum_{k} \left( a^{\dagger}_k g^{\dagger} f 
        + \text{h.c.} \right).
\end{align}

The Hamiltonian \eqref{eq:model:hamiltonian_realspace} exhibits strong similarities to 
well-known models in quantum optics and condensed-matter theory. Specifically, the 
Hamiltonian could either be seen as a spin-boson model, a multi-mode generalization of the Jaynes-Cummings 
model 
\cite{JaynesCummings-1963}, 
a Dicke model 
\cite{Dicke-1954}, 
or a spinless, partly bosonic version of the Anderson impurity model
\cite{Anderson-1961} 
without on-site interaction. As discussed above, this invites the adaptation of tools
that have been developed for the analyses of these models, albeit bearing in mind the
different physics and the resulting different questions that shall be addressed. In
fact, path-integral and related Green's function techniques are among the most flexible
approaches, allowing for closed-form solutions in simple cases and providing efficient
perturbative approaches for challenging cases such as disordered systems\cite{Rammer04} and systems 
out of equilibrium\cite{Rammer07}. 

\section{Path integral approach}
\label{sec:pathint}
We begin our exposition by defining the Green's function as the matrix element of an 
initial and a final state at different times
\begin{equation}\label{eq:pathint:G_matrixelement}
	G(t_f-t_i) = -i \langle f | e^{- \int_{t_i}^{t_f} H(t) dt } | i \rangle,
\end{equation}
where $H$ is a generic Hamiltonian. In addition, $|i\rangle$ and $|f\rangle$ represent,
respectively, the initial and the final state. \\
For a path-integral representation, we utilize the resolution of unity via coherent states 
\begin{equation}\label{eq:pathint:CoherentUnity}
	\mathbb{1} = \frac{1}{(2\pi i)^c}
	             \int \prod_{\alpha} \mathrm{d} \psi_{\alpha}\mathrm{d}\psi^*_{\alpha} 
	             e^{-\sum_{\alpha}\psi^{*}_{\alpha}\psi_{\alpha}}
	             \vert \psi_{\alpha}\rangle\langle \psi_{\alpha} \vert,
\end{equation}
where $c=1$ for complex fields, $c=0$ for Grassman fields, and $\alpha$ labels the set 
of associated one-particle states ($\vert \psi_{\alpha}\rangle\langle \psi_{\alpha} \vert$
being the corresponding projection operator). Inserting this resolution of unity twice into 
Eq.~\eqref{eq:pathint:G_matrixelement} yields
\begin{widetext}

\begin{equation}\label{eq:pathint:G_coherentstates}
	G(t_f-t_i) = -i \int \prod_{m,n} \mathrm{d}\psi_{i,m} \mathrm{d}\psi^*_{i,m} \mathrm{d}\psi_{f,n} \mathrm{d}\psi^*_{f,n}  e^{ -\sum_{m}\psi^{*}_{i,m} \psi_{i,m} -\sum_{n}\psi^{*}_{f,n} \psi_{f,n}}  \langle f | \psi_{f} \rangle  \langle \psi_{i} | i \rangle 
                  \, G(f,i;t_f-t_i).
\end{equation}
Here $G(f,i,t_f-t_i)$ is given by
\begin{equation}
	G(f,i;t_f-t_i) = \langle \psi_{f} | e^{- \int_{t_i}^{t_f} H(t) dt } | \psi_{i} \rangle,
\end{equation}
so that the labels $i$ and $f$ represent the initial and final fields, $\psi_i$ and $\psi_f$, 
respectively. In the above equations, the indices $m$ and $n$ run over the number of different 
modes that occur in the generic Hamiltonian $H$ by virtue of corresponding creation and annihilation 
operators.

In the case of the Hamiltonian given by Eq.~\eqref{eq:model:hamiltonian}, we have to introduce 
three types fields: Complex fields $\phi_k$ for the bosonic modes in the waveguide and two 
Grassmann fields $\theta$ and $\chi$, which correspond to the fermionic operators $f$ and $g$, 
respectively. With these definitions, Eq.~\eqref{eq:pathint:G_coherentstates} reads

\begin{equation}\label{eq:pathint:G_bosonicIntegration}
 G(t_f-t_i) = -i \int \prod_{k,k'} \mathrm{d}\phi_{i,k} \mathrm{d}\phi^*_{i,k}  
	                                  \mathrm{d}\phi_{f,k'} \mathrm{d}\phi^*_{f,k'}
                 e^{- \sum_{k}\phi^{*}_{i,k} \phi_{i,k} 
                            - \sum_{k}\phi^{*}_{f,k} \phi_{f,k}} 
                 \langle \text{ph}_f | \phi_{f} \rangle  \langle \phi_{i} | \text{ph}_i \rangle 
                  G(f,i;t_f-t_i),
\end{equation}
where we have absorbed the integration over the Grassman fields into

\begin{align}\label{eq:pathint:Gfi_fermions}
 G(f,i;t_f-t_i) = & \int \mathrm{d}\chi_{i} \mathrm{d}\chi^*_{i}  
                          \mathrm{d}\chi_{f} \mathrm{d}\chi^*_{f} 
                    \mathrm{d}\theta_{i} \mathrm{d}\theta^*_{i}  
                            \mathrm{d}\theta_{f} \mathrm{d}\theta^*_{f} 
                    e^{ -\chi^{*}_{i} \chi_{i} - \chi^{*}_{f} \chi_{f} 
                                -\theta^{*}_{i} \theta_{i} - \theta^{*}_{f} \theta_{f}} \nonumber \\
                  & \times \langle \text{TLS}_f | \chi_{f} \theta_{f} \rangle  
                            \langle \chi_{i} \theta_{i} | \text{TLS}_i \rangle  
                    \langle \phi_{f} \chi_{f} \theta_{f} 
                            \vert e^{- \int_{t_i}^{t_f} H(t) dt } \vert 
                            \phi_{i} \chi_{i} \theta_{i} \rangle.
\end{align}
In the above equations, we have employed a decomposition of the system's initial state 
into a product of a part for the photons with a part for the two-level system
\begin{equation}
	\vert i \rangle = \vert \text{ph}_{i} \rangle \otimes \vert \text{TLS}_{i} \rangle.
\end{equation}
In addition, we have employed an analogous decomposition for the final state.

For the calculation of $G(f,i,t_f-t_i)$ we mainly follow the lines of Ref. 
\onlinecite{BoudjedaaBounamesNouicerChetouaniHammann_1996}, 
so that we will restrict ourselves to the relevant intermediate steps. Upon inserting the 
resolution of unity $N$ times in Eq.~\eqref{eq:pathint:Gfi_fermions}, 
we arrive at
\begin{align} \label{eq:pathint:DefTransAmpl}
	G(f,i;t_f-t_i)=& \int \mathrm{d}\chi_{i} \mathrm{d}\chi^*_{i}  
                          \mathrm{d}\chi_{f} \mathrm{d}\chi^*_{f} 
                    \mathrm{d}\theta_{i} \mathrm{d}\theta^*_{i}  
                            \mathrm{d}\theta_{f} \mathrm{d}\theta^*_{f} 
                    e^{ -\chi^{*}_{i} \chi_{i} - \chi^{*}_{f} \chi_{f} 
                                -\theta^{*}_{i} \theta_{i} - \theta^{*}_{f} \theta_{f}}  \langle \text{TLS}_f \vert \chi_{f} \theta_{f} \rangle  
                                        \langle \chi_{i} \theta_{i} \vert \text{TLS}_i \rangle \nonumber \\ 
                  & \times \lim_{N \rightarrow \infty} \int \prod_{m=1}^{N} \prod_{n=1}^{N-1} \prod_{k}               
                      \frac{\mathrm{d}\phi_{n,k} \mathrm{d}\phi^*_{n,k}
                            \mathrm{d}\theta_n   \mathrm{d}\theta_n^*
                            \mathrm{d}\chi_n \mathrm{d}\chi^*_n}{2\pi i}  \nonumber \\
                  &  \times e^{-\sum_{n,k} ( \phi_{n,k}\phi^*_{n,k}+\chi_{n}\chi^*_{n}+\theta_{n}\theta^*_{n} )}
                   e^{-i\eta H( \phi_{m},\phi^*_{m-1},\chi_{m},\chi^*_{m-1},\theta_{m},\theta^*_{m-1})},
\end{align}
where $\eta = \frac{t_f-t_i}{N}$ and
\begin{align}\label{eq:pathint:HamiltonianCoherent}
	H( \phi_{n},&\phi^*_{n-1},\chi_{n},\chi^*_{n-1},\theta_{n},\theta^*_{n-1}) = \nonumber \\
     &  \sum_k \epsilon(k) \phi_{n,k}\phi^*_{n-1,k}
       + \frac{\Omega}{2}(\theta_{n}\theta^*_{n-1}-\chi_{n}\chi^*_{n-1})+ 
      \frac{U}{\sqrt{N}}
       \sum_k \left( \phi^*_{n,k}\chi^*_{n}\theta_{n-1}+\theta^*_{n}\chi_{n-1}\phi_{n-1,k}\right).
\end{align}
The labels $n=0$ and $n=N$ correspond, respectively, to the initial and final fields. Next, we 
integrate out the intermediate fermionic degrees of freedom and obtain
\begin{align} \label{eq:pathint:IntOutFerm}
	G(f,i;t_f-t_i) = & \int \mathrm{d}\chi_{i} \mathrm{d}\chi^*_{i}  
                          \mathrm{d}\chi_{f} \mathrm{d}\chi^*_{f} 
                    \mathrm{d}\theta_{i} \mathrm{d}\theta^*_{i}  
                            \mathrm{d}\theta_{f} \mathrm{d}\theta^*_{f} 
                    e^{ -\chi^{*}_{i} \chi_{i} - \chi^{*}_{f} \chi_{f} 
                                -\theta^{*}_{i} \theta_{i} - \theta^{*}_{f} \theta_{f}}  \langle \text{TLS}_f \vert \chi_{f} \theta_{f} \rangle  
                                        \langle \chi_{i} \theta_{i} \vert \text{TLS}_i \rangle \nonumber \\ 
	          & \times \lim_{N \rightarrow \infty} \int\prod_{n=1}^{N-1}\prod_k
	                  \frac{\mathrm{d}\phi_{n,k}\mathrm{d}\phi^*_{n,k}}{2\pi i} 
	               e^{-\sum_{n,k} \phi_{n,k}\phi^*_{n,k}(1-i\eta\epsilon(k))}                  
                           e^{\vec{q}^{\dag}_N R(\phi,\phi^*)\vec{q}_0}.
\end{align}
Here, we have introduced the following abbreviations

\begin{equation}
 q_i  = (\theta_i, \chi_i), \qquad R(\phi,\phi^*)  = R(\phi_N,\phi_{N-1}^*)\cdot...\cdot R(\phi_1,\phi_0^*), \nonumber 
\end{equation}
\begin{equation}\label{eq:pathint:Abbreviation1}
 	R(\phi_i,\phi_{i-1}^*)  = \left( \begin{array}{cc} 
	                                     1-i\eta\Omega/2 & 0 \\
                                                     0 & 1+i\eta\Omega/2\\
                                    \end{array} \right) 
                          - i \eta \frac{U}{\sqrt{L}} \sum_k 
                             \left( \begin{array}{cc}            
                                                  0 & \phi_{i-1,k}\\
                                       \phi^*_{i,k} & 0\\
                                    \end{array} \right). 
\end{equation}

The TLS can either be in the excited or in the ground state. Thus, we can write the TLS' state 
as a vector
\begin{equation}
	\vert TLS \rangle =  \begin{pmatrix}
                          \vert e \rangle \\
                          \vert g \rangle 
                       \end{pmatrix},
\end{equation}
which induces a matrix structure to $G(f,i;t_f-t_i)$
\begin{equation}\label{eq:pathint:G_MatrixStructure}
	G(f,i;t_f-t_i) = \begin{pmatrix}
                      G_{\mathrm{e}} (f,i;t_f-t_i) & G_{\mathrm{ab}} (f,i;t_f-t_i) \\
		                    G_{\mathrm{em}} (f,i;t_f-t_i) & G_{\mathrm{w}} (f,i;t_f-t_i)
                  \end{pmatrix}.
\end{equation}
The nomenclature of this matrix notation is as follows. 
We denote the case when the TLS is excited in the initial as well as in the final state by the 
diagonal element $G_{\mathrm{e}}$. Consequently, $G_{\mathrm{e}}$ covers the dynamics of the quantum impurity, 
i.e., the TLS that interacts with photons from the waveguide so that we call it the TLS-Green's 
function. 
Similarly, we denote the case when the TLS is in the ground state for both the initial and the final 
state by the diagonal element $G_{\mathrm{w}}$. Clearly, this quantity describes the dynamics of the photons 
in the waveguide in the presence of the TLS and we, therefore, call it the waveguide Green's function.
The off-diagonal element $G_{\mathrm{ab}}$ features that the TLS is initially in the ground state, but ends 
up being in the excited state. This means that the TLS has absorbed a photon, hence we call $G_{\mathrm{ab}}$ 
the absorption Green's function. Clearly, $G_{\mathrm{em}}$ covers the complementary process and we call it 
the emission Green's function.

We now utilize the fermionic resolution of unity and the identities
\cite{BoudjedaaBounamesNouicerChetouaniHammann_1996} 
$\langle e \vert \theta \chi \rangle = \theta$, $ \langle g \vert \theta \chi \rangle = \chi$, and 
$\theta \chi^*  = e^{-\theta \chi^*}-1$ to perform the last integration over the fermionic fields. 
We will restrict ourselves to the TLS-Green's function, the other Green's functions can be determined 
in exactly the same way. Following the procedure in Ref. \onlinecite{BoudjedaaBounamesNouicerChetouaniHammann_1996}, 
we find

\begin{equation}\label{eq:pathint:Projection}
G_{\mathrm{e}}(f,i,t_f-t_i) =  \lim_{N \rightarrow \infty} \int \prod_{n=1}^{N-1}\prod_k 
                                                    \frac{\mathrm{d}\phi_{k,n} \mathrm{d}\phi^*_{k,n}}{2\pi i} 
                 e^{-\sum_{n,k} \phi_{n,k}\phi^*_{n,k}(1-i\eta\epsilon(k))} R(\phi, \phi^*).
\end{equation}
In the next step, we want to integrate out the bosonic fields. Therefore, we use the matrix 
$R(\phi,\phi^*)$ (see Eq.~\eqref{eq:pathint:Abbreviation1}) and expand it up to $\mathcal{O}(\eta)$. Taking the limit $N\rightarrow\infty$ 
at the end, we find
\begin{align}\label{eq:pathint:Bosons}
	G_{\mathrm{e}}(f,i;t_f-t_i)  = 
	& \sum_{r=0}^{\infty} \left(\frac{iU}{\sqrt{L}}\right)^r 
	  \int_0^{t_i-t_f} \mathrm{d}t_{2r} \int_0^{t_{2r}} \mathrm{d}t_{2r-1}....\int_0^{t_{2}}\mathrm{d}t_{1}
	  \prod_{k} \prod_{m=1}^{2r} \frac{\mathrm{d} \phi_{m,k} \mathrm{d} \phi^*_{m,k}}{2\pi i}\nonumber \\
	& \times G_{\mathrm{e},0}(\phi^*_N,\phi_{2r},t_f-t_i-t_{2r}) \,
	         e^{-\sum_{k}\phi^*_{2r,k}\phi_{2r,k}}
	         \sum_k\phi_{2r,k}\nonumber \\
	& \times G_{\mathrm{w},0}(\phi^*_{2r},\phi_{2r-1},t_{2r}-t_{2r-1}) \,
	         e^{-\sum_{k}\phi^*_{2r-1,k}\phi_{2r-1,k}}
	         \sum_k\phi^*_{2r-1,k}\nonumber \\
	& \times G_{\mathrm{e},0}(\phi^*_{2r-1},\phi_{2r-2},t_{2r-1}-t_{2r-2}) \,
	         e^{-\sum_{k}\phi^*_{2r-2,k}\phi_{2r-2,k}}
	         \sum_k\phi_{2r-2,k}\nonumber \\
  & ...........................\nonumber \\
  &\times G_{\mathrm{e},0}(\phi^*_{1},\phi_{0},t_1).
\end{align}
Here, we have introduced the free propagators of the excited TLS and the waveguide 
\begin{equation}
	G_{\mathrm{e}/w,0}(\phi^*_i,\phi_j,t) = e^{\mp i\frac{\Omega}{2}t} \, e^{\sum_k\phi^*_{i,k} e^{-i\epsilon(k)t}\phi_{j,k}}.
\end{equation}
Further, we would like to note that at this point the correspondence to an ordinary perturbation 
series becomes clear. We have vertices with strength $\frac{U}{\sqrt{L}}$ at which a photon is 
annihilated or created and the free propagation of photons or excitations between two successive
scattering events. 

In order to calculate the full Green's function, we have to evaluate Eq.~\eqref{eq:pathint:G_bosonicIntegration}. 
For an $n$-photon state, the projection on the coherent states is given by
\begin{equation}\label{eq:pathint:BosonsProjection}
	\langle k_1 k_2 \dots k_n \vert \phi \rangle = \phi_{k_1} \phi_{k_2} \dots \phi_{k_n}.
\end{equation}
Inserting these bosonic fields and integrating out the bosonic variables finally 
yields the full Green's functions. As our basic (and equivalent) Hamiltonians, Eqs. 
\eqref{eq:model:hamiltonian_realspace} and \eqref{eq:model:hamiltonian_momentumspace}, 
conserve the total number of excitations (cf. Eq.~\eqref{eq:model:excitationnumber}), 
we proceed by evaluating the TLS-Green's function explicitly for the single- and 
two-excitation sectors.

\subsection{Single excitation sector}
\label{sec:pathint:spcase}
For a single excitation, we start the evaluation of the TLS-Green's function ${_{1}G_{\mathrm{e}}(t_f-t_i)}$ 
by noting that we have an initial and final state where the TLS is excited and there are no photons 
in the waveguide. The TLS-Green's function then reads

\begin{equation}\label{eq:pathint:spcase:IntOutBosonsI}
	 {_{1}G_{\mathrm{e}}(t_f-t_i)}  =  -i \int \prod_{k,k'} \mathrm{d}\phi_{f,k} \mathrm{d}\phi^*_{f,k} 
	                                                              \mathrm{d}\phi_{i,k'}\mathrm{d}\phi^*_{i,k'}         
	                                          e^{ -\sum_{k} \phi_{f,k}\phi^*_{f,k} 
	                                                       -\sum_{k} \phi_{i,k}\phi^*_{i,k}} 
                                             G_{\mathrm{e}}(f,i,t_f-t_i).
\end{equation}
The dependence on the fields $\phi_f, \phi_i$ is easily integrated out and we obtain
\begin{align}\label{eq:pathint:spcase:Gd_pert_time}
	 {_{1}G_{\mathrm{e}}(t_f-t_i)}  = & -i \sum_{r=0}^{\infty} \left(\frac{iU}{\sqrt{L}}\right)^{2r} \int_0^{t_i-t_f}\mathrm{d}t_{2r}\int_0^{t_{2r}}\mathrm{d}t_{2r-1}....\int_0^{t_{2}}\mathrm{d}t_{1}\nonumber \\
					  & \times G^0_{\mathrm{e}}(t_f-t_i-t_{2r}) G^{0,\Sigma}_{\mathrm{w}}(t_{2r}-t_{2r-1}) G^0_{\mathrm{e}}(t_{2r-1}-t_{2r-2}) \dots G^0_{\mathrm{e}}(t_{1}-t_{0}) \nonumber \\
					  = & \sum_{r} {_{1}G^{(r)}_{\mathrm{e}}(t_f-t_i)} 
\end{align}
We note that every term of the sum has the form of a Dyson series and represents a convolution 
\mbox{${_{1}G_{\mathrm{e}}^{(r)}} \sim \underbrace{G^0_{\mathrm{e}}\ast G^{0,\Sigma}_{\mathrm{w}}\ast G^0_{\mathrm{e}}....\ast G^0_{\mathrm{e}}}_{\text{$2r+1$ factors}}$ }
with 
\begin{equation}\label{eq:pathint:spcase:G0d}
	{_{1}G^0_{\mathrm{e}}(t)} = e^{-i\frac{\Omega}{2}t},\qquad {_{1}G^{0,\Sigma}_{\mathrm{w}}(t)} = e^{i\frac{\Omega}{2}t}\sum_k e^{- i \epsilon (k)t} = \sum_k {_{1}G^{0}_{\mathrm{w}}(k,t)}.
\end{equation}
By means of the convolution theorem, we can recast this in Fourier space as an algebraic 
equation 
\begin{equation}\label{eq:pathint:spcase:IntOutBosonsIII}
	 {_{1}G_{\mathrm{e}}(\omega)}  = \sum_{r=0}^{\infty} \frac{1}{\omega-\Omega/2+i\delta}  \left(\frac{1}{\omega-\Omega/2+i\delta}\sum_k\frac{U^2/L}{\omega+(\Omega/2-\epsilon(k))+i\delta}\right)^r.
\end{equation}
Here, we have introduced the factors $+i \delta$ because we are working with retarded 
Green's functions, i.e., the limit $\delta \to 0+$ should be understood.\\
The above equation can readily be solved, and we arrive at the main equation of this section,
the TLS-Green's function ${_{1}G_{\mathrm{e}}(\omega)}$ in the single-excitation sector
\begin{equation}\label{eq:pathint:spcase:Gd}
	{_{1}G_{\mathrm{e}}(\omega)}  = \frac{1}{\omega-\Omega/2+i\delta- {_{1}\Sigma(\omega)}},
\end{equation}
where the self-energy ${_{1}\Sigma(\omega)}$ is given by
\begin{equation}\label{eq:pathint:spcase:Sigma}
 {_{1}\Sigma(\omega)} = \frac{U^2}{L} \sum_{k} \frac{1}{\omega +\Omega/2 - \epsilon(k) + i \delta}.
\end{equation}
The waveguide-Green's function in the single excitation sector can be calculated in a similar 
manner as the above TLS-Green's function. The main difference lies in the fact that the initial 
and final states feature a free photon. As a result, one has to add free photon fields in Eq. 
\eqref{eq:pathint:spcase:IntOutBosonsI}, which yields
\begin{equation}\label{eq:pathint:spcase:IntOutBosonsI_wire}
	 {_{1}G_{\mathrm{w}}}  (k_i,k_f;t_f-t_i)  = -i \int \prod_{k,k'} \mathrm{d}\phi_{f,k}  \mathrm{d}\phi^*_{f,k}  
							\mathrm{d}\phi_{i,k'} \mathrm{d}\phi^*_{i,k'} \phi^{*}_{i,k_i} \phi_{f,k_f} 
							e^{-\sum_{k} \phi_{f,k}\phi^*_{f,k} -\sum_{k} \phi_{i,k}\phi^*_{i,k}}  G_{\mathrm{w}}(f,i;t_f-t_i).
\end{equation}
Performing the integrations and again transiting to momentum space results in
\begin{equation}\label{eq:pathint:spcase:Gwfull_w}
 {_{1}G_{\mathrm{w}}(k_i,k_f;\omega)} =  {_{1}G^{0}_{\mathrm{w}}}(k_i;\omega) \delta_{k_i,k_f}  + 
					\frac{U^2}{L} \,{_{1}G^{0}_{\mathrm{w}}}(k_i;\omega) \,{_{1}G_{\mathrm{e}}}(\omega) \,{_{1}G^{0}_{\mathrm{w}}}(k_f;\omega),
\end{equation}
where
\begin{equation}\label{eq:pathint:spcase:Gw0_w}
 {_{1}G^{0}_{\mathrm{w}}(k;\omega)} = \frac{1}{\omega + \Omega/2 - \epsilon(k) + i\delta}.
\end{equation}
The expression \eqref{eq:pathint:spcase:Gwfull_w} for the waveguide-Green's function
${_{1}G_{\mathrm{w}}(k_i,k_f;\omega)}$ consists two terms, which can be identified with free 
propagation of the photon and scattering off the (renormalized) TLS, respectively. 
The corresponding absorption and emission Green's functions are derived in Appendix 
\ref{sec:appendix:spGF}, together with their diagrammatic representation.

\subsection{Two excitation sector}
\label{sec:pathint:2pcase}
In this sub-section, we consider two excitations in our system. Again, we start with 
determining the Green's function for an excited impurity in the initial and in the 
final state, i.e., the TLS-Green's function $_{2}G_{\mathrm{e}}$. The starting point is the 
general expression given by Eq.~\eqref{eq:pathint:G_bosonicIntegration}. As compared
with the single-excitation case, the only difference is that, according to Eq. 
\eqref{eq:pathint:BosonsProjection},  we now have an additional bosonic excitation 
in the in- and out state. Consequently, for this specific case $G(t_f-t_i)$ reads 
\begin{equation}\label{eq:pathint:2pcase:IntOutBosonsI}
 {_{2}G_{\mathrm{e}}} (k_i,k_f;t_f-t_i) = -i \int \prod_{k,k'} \mathrm{d}\phi_{f,k} \mathrm{d}\phi^*_{f,k} 
                                                    \mathrm{d}\phi_{i,k'}\mathrm{d}\phi^*_{i,k'}  \phi^{*}_{i,k_i} \phi_{f,k_f} \, 
	                                      e^{-\sum_{k} \phi_{f,k}\phi^*_{f,k} -\sum_{k} \phi_{i,k}\phi^*_{i,k}}  G_{\mathrm{e}}(f,i,t_f-t_i),
\end{equation}
where $G_{\mathrm{e}}(f,i,t_f-t_i)$ is given by \eqref{eq:pathint:Bosons}. Integration over 
the bosonic degrees of freedom is performed along the same lines as in the previous 
sub-section. Again transiting to momentum space, the Green's function is given by
\begin{align}\label{eq:pathint:2pcase:Gd_pertubationseries}
  _{2} G_{\mathrm{e}}(k_f,k_i;\omega) = &{_{2}G_{\mathrm{e},\mathrm{r}}}(k_i;\omega) \delta_{k_i,k_f} \nonumber \\
                                        &+ \frac{U^2}{L} \,{_{2}G_{\mathrm{e},\mathrm{r}}}(k_i;\omega)  
							 \,{_{2}G^{0}_{\mathrm{w}}}(k_f,k_i;\omega)
                                                         \,{_{2}G_{\mathrm{e},\mathrm{r}}}(k_f;\omega) \nonumber \\
                                        &+ \frac{U^4}{L^2} \sum_k 
                                            \,{_{2}G_{\mathrm{e},\mathrm{r}}}(k_i;\omega)  
					    \,{_{2}G^{0}_{\mathrm{w}}}(k_i,k;\omega)  
					    \,{_{2}G_{\mathrm{e},\mathrm{r}}}(k;\omega)   
					    \,{_{2}G^{0}_{\mathrm{w}}}(k,k_f;\omega)
                                            \,{_{2}G_{\mathrm{e},\mathrm{r}}}(k_f;\omega) \nonumber \\
					& + \dots ,
\end{align}
where
\begin{equation}\label{eq:pathint:2pcase:Gw0}
	_{2}G^{0}_{\mathrm{w}}(k,k';\omega)=\frac{1}{\omega+\Omega/2-\epsilon(k)-\epsilon(k')+i\delta}
\end{equation}
and
\begin{equation}\label{eq:pathint:2pcase:Gdr}
	_{2}G_{\mathrm{e},\mathrm{r}}(k;\omega)  = \frac{1}{\omega-\Omega/2-\epsilon(k)+i\delta- {_{2}\Sigma}(k;\omega)}.
\end{equation}
The self-energy ${_{2}\Sigma}(k;\omega)$ is given by
\begin{equation}\label{eq:pathint:2pcase:Selfenergy}
	{_{2}\Sigma}(k;\omega) = \frac{U^2}{L} \sum_{k'}  {_{2}G^{0}_{\mathrm{w}}} (k,k';\omega) .
\end{equation}

While we cannot express ${_{2}G_{\mathrm{e}}}(k_f,k_i;\omega)$ in closed form, we can recast
Eq.~\eqref{eq:pathint:2pcase:Gd_pertubationseries} in a T-Matrix representation which
can easily be accessed numerically. Explicitly, the T-Matrix representation reads 
\begin{equation}\label{eq:pathint:2pcase:Gd_Tmatrix}
	{_{2}G_{\mathrm{e}}}(k_f,k_i;\omega)  = {_{2}G_{\mathrm{e},\mathrm{r}}}(k_i;\omega)\delta_{k_i,k_f}
						+ \, {_{2}G_{\mathrm{e},\mathrm{r}}}(k_i;\omega)  \,T(k_f,k_i;\omega) 
						  \, {_{2}G_{\mathrm{e},\mathrm{r}}}(k_f;\omega) 
\end{equation}
with the T-matrix
\begin{equation}\label{eq:pathint:2pcase:TMatrix}
	T(k_f,k_i;\omega) = \frac{U^2}{L} {_{2}G^{0}_{\mathrm{w}}}(k_f,k_i;\omega)  
				+ \frac{U^2}{L} \sum_{k} \,{_{2}G^{0}_{\mathrm{w}}}(k,k_i;\omega)  
                                                 \,{_{2}G_{\mathrm{e},\mathrm{r}}}(k;\omega)  
                                                \,T(k_f,k;\omega).
\end{equation}
The above T-matrix representation of ${_{2}G_{\mathrm{e}}}(k_f,k_i;\omega)$, Eqs. \eqref{eq:pathint:2pcase:Gd_Tmatrix}
and \eqref{eq:pathint:2pcase:TMatrix}, constitutes one of the main results of our work. 
Eq.~\eqref{eq:pathint:2pcase:Gd_Tmatrix} describes the nontrivial behavior of the TLS 
in the presence of an additional photon. Clearly, the result is more complicated than 
in the single-excitation case, where we were able to solve a standard Dyson equation. 
In the present case of two excitations, however, we have found a self-consistent 
description of the Green's function, which can be solved numerically for arbitrary
dispersion relations. Furthermore, in the special case of a linear dispersion relation, 
the TLS-Green's function can even be calculated analytically (see section \ref{sec:2pcase:Linear}).

The corresponding waveguide-Green's function, which describes the behavior of two photons 
in the presence of an impurity in the ground state is given by
\begin{align}\label{eq:pathint:2pcase:Gw_InOutBosons}
 {_{2}G_{\mathrm{w}}}(k_i,p_i,k_f,p_f;t_f-t_i) =   -i \int \prod_{k,k'} &\mathrm{d}\phi_{f,k} \mathrm{d}\phi^*_{f,k} 
					\mathrm{d}\phi_{i,k'}\mathrm{d}\phi^*_{i,k'} \nonumber \\   
					&\times \phi^{*}_{i,k_i} \phi^{*}_{i,p_i} \phi_{f,k_f} \phi_{f,p_f} \, 
					e^{-\sum_{k} \phi_{f,k}\phi^*_{f,k} 
					-\sum_{k} \phi_{i,k}\phi^*_{i,k}} 
					G_{\mathrm{w}}(f,i,t_f-t_i).
\end{align}
After performing the integrations, the waveguide-Green's function finally reads
\begin{align}\label{eq:pathint:2pcase:Gw_result}
 _{2}G_{\mathrm{w}}(\omega,k_f,p_f,k_i,p_i) = &{_{2}G^0_{\mathrm{w}}}(k_i,p_i;\omega) 
						\left( \delta_{k_i,k_f}\delta_{p_i,p_f} + 
						\delta_{k_i,p_f}\delta_{p_i,k_f} \right) \nonumber\\
					      &+\frac{U^2}{L} \,{_{2}G^0_{\mathrm{w}}}(\omega,k_i,p_i)  
                                                    \,{_{2}G^{\text{sym}}_{\mathrm{e}}}(k_i,p_i,k_f,p_f;\omega) 
                                                    \,{_{2}G^0_{\mathrm{w}}}(\omega,k_f,p_f) ,
\end{align}
where ${_{2}G^{\text{sym}}_{\mathrm{e}}}$ is a symmetrized version of the full TLS-Green's function and 
is given by
\begin{equation}\label{eq:pathint:2pcase:Gd_sym}
 {_{2}G^{\text{sym}}_{\mathrm{e}}}(k_i,p_i,k_f,p_f;\omega) =  {_{2}G_{\mathrm{e}}}(k_f,k_i;\omega) + {_{2}G_{\mathrm{e}}}(k_f,p_i;\omega)  + {_{2}G_{\mathrm{e}}}(p_f,k_i;\omega) + {_{2}G_{\mathrm{e}}}(p_f,p_i;\omega).
\end{equation}
\end{widetext}
For the case of two excitations, the full waveguide-Green's function ${_{2}G_{\mathrm{w}}(\omega,k_f,p_f,k_i,p_i)}$ 
exhibits the same structure as in the single-excitation sector. It consists of a free propagation term 
and a second term that describes the scattering off a renormalized TLS. For the absorption and emission
Green's functions, we provide the expressions for the two-excitation sector in Appendix 
\ref{sec:appendix:2pGF} along with their diagrammatic representation.

\section{Feynman Diagram representation}
\label{sec:FeynmanDiagrams}

\begin{table}[b!]
 \begin{center}
  \begin{tabular}{c|c|c|c}
   Photons & $| e \rangle $ & $| g \rangle$ & Vertex \\ \hline \hline
   \showgraph{width=1.8cm}{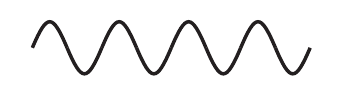} & \showgraph{width=1.8cm}{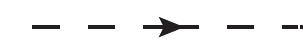} & \showgraph{width=1.8cm}{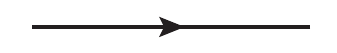} & \showgraph{width=1.8cm}{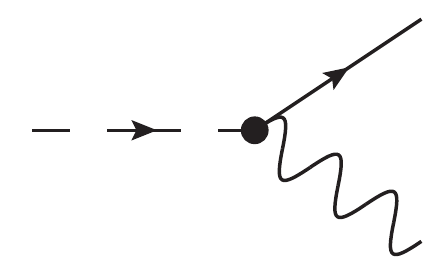}
  \end{tabular}
 \end{center}
 \caption{Table of the representation of the individual species and the interaction vertex in terms of Feynman diagrams.}
 \label{tab:FeynmanPresentations}
\end{table}

In this section, we illustrate the formulas obtained by the path integral approach by way of Feynman 
diagrams. Specifically, we will refrain from providing a rigorous derivation of the full diagrammatic 
technique but will instead represent the equations of the previous section by Feynman diagrams. 
This "visualization" provides a clear identification and interpretation of physical processes and, 
as already alluded to above, facilitates a very flexible and efficient framework for perturbative analyses.\\
In general, the Hamiltonian given by Eq.~\eqref{eq:model:hamiltonian} features three distinct species 
of quantized excitations, i.e., bosons (photons) with a mode index $k$ and two types of fermions, representing the excited 
and the ground state of the TLS. 
In the following diagrammatic representation, these excitations will be 
depicted by a wavy, a dashed and a solid line, respectively. This mapping is also shown in Tab.~\ref{tab:FeynmanPresentations} 
for clarification. Apart from these diagonal contributions, Eq.~\eqref{eq:model:hamiltonian} also features a scattering 
vertex, which connects the individual lines and is also shown in Tab.~\ref{tab:FeynmanPresentations}. 

As much of the physical insight to be gained originates from a comparison of the results for the
case of a single excitations with the case of two excitations, we will proceed in a corresponding
sequence of sub-sections.

\subsection{Single excitation sector}
\label{sec:FeynmanDiagrams:spcase}
In the case of a single excitation our first goal is to depict the TLS-Green's function $_{1}G_{\mathrm{e}}$ 
in terms of Feynman diagrams. Upon inspecting Eq.~\eqref{eq:pathint:spcase:Gd_pert_time} we find two distinct 
contributions. The first contribution stems from

\begin{equation}
 {_{1}G^{0}_{\mathrm{e}}}(t'-t) = e^{-i \frac{\Omega}{2} (t'-t)} = \showgraph{width=2cm}{GFgeneral_f.pdf}
\end{equation}
and describes the propagation of an excited TLS from time $t$ to $t'$. The second contribution is 
given by (cf. Eq.~\eqref{eq:pathint:spcase:G0d})

\begin{equation}\label{eq:FeynmanDiagrams:spcase:Gwsigma_t}
 {_{1}G^{0}_{\mathrm{w}}}(k,t' - t) = e^{i \frac{\Omega}{2} (t' - t)} e^{-i \epsilon(k) (t' - t)}  = \showgraph{width=2cm}{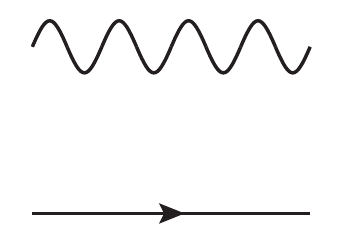},
\end{equation}
and describes the simultaneous transport of a TLS in the ground state together with a photon in 
the waveguide from time $t$ to $t'$. Furthermore, we can infer from Eq.~\eqref{eq:pathint:spcase:Gd_pert_time} 
that each vertex is weighted by a factor $iU/\sqrt{L}$. Combining everything we can rewrite the TLS-Green's 
function diagrammatically as

\begin{align}\label{eq:FeynmanDiagrams:spcase:Gd_pertubationseries}
 {_{1}G_{\mathrm{e}}}(t_f-t_i) =-i \Bigl\{ & \showgraph{width=1.5cm}{GFgeneral_f.pdf} \nonumber \\
                                          & + \showgraph{width=3cm}{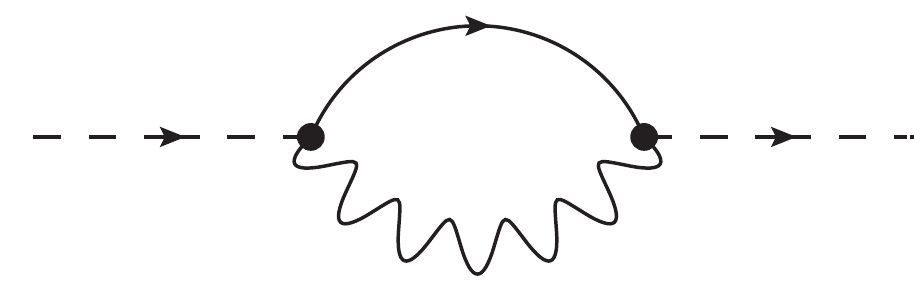} \nonumber \\
                                          & + \showgraph{width=4.5cm}{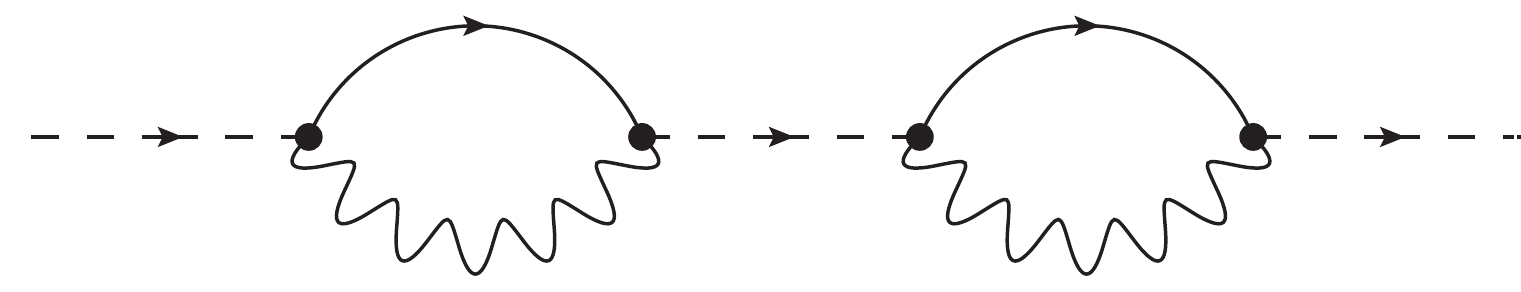} \nonumber \\
                                          & + \dots \Bigr\}
\end{align}
where each vertex provides an integration over internal times. Additionally, each photonic line 
sandwiched between two interaction vertices implies a summation over the corresponding momentum. 
At this point we are able to exploit one of the main advantages of the diagrammatic approach and
provide an interpretation of the the individual terms in the perturbation series in terms of 
physical processes: The excited TLS goes to the ground state by emitting a photon. This photon 
is absorbed  at a later time, setting the TLS to the excited state again. This process is 
repeated $n$ times in the $n$th term of the perturbation series.

In the frequency domain, the convolution integrals associated with the time-domain diagrams 
turn into simple products, so that we may retain the free propagation and the bubble diagram
as well as the entire perturbation series. Therefore, in the frequency domain no integration
with regards to the scattering vertices are implied and per bubble only a weighting factor of
$U/\sqrt{L}$ is implied, furthermore the global prefactor $-i$ can be omitted. However, each 
sandwiched photonic line still comes with a summation over the corresponding momentum. Explicitly, 
the free TLS-Green's function reads
\begin{equation}
 \showgraph{width=2cm}{GFgeneral_f.pdf} = \frac{1}{\omega - \Omega/2 + i \delta} 
                                            = {_{1}G^{0}_{\mathrm{e}}}(\omega).
\end{equation}
Similarly, in the frequency domain, we evaluate the bubble diagram after cutting it free from the interaction vertices to
\begin{align}
 \showgraph{width=2cm}{1Gw0.pdf} & =  \frac{1}{\omega + \Omega/2 - \epsilon(k) + i\delta} \nonumber \\
                                          & =  {_{1}G^{0}_{\mathrm{w}}}(k;\omega).
\end{align}
In the time-domain this Green's function describes the simultaneous propagation of a TLS in 
the ground state and a photon in the waveguide. Since both excitations are created and annihilated 
at the same times, the Fourier transform yields only one frequency argument and leads to the 
analytic form shown above.

As usual, we can cast the perturbation series \eqref{eq:FeynmanDiagrams:spcase:Gd_pertubationseries} 
into the form of a self-consistent Dyson equation
\begin{align}
 {_{1}G_{\mathrm{e}}}(\omega)  & = \showgraph{width=2cm}{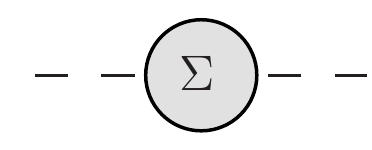} \nonumber \\
                                        & = \showgraph{width=1cm}{GFgeneral_f.pdf} 
                                            + \showgraph{width=3.5cm}{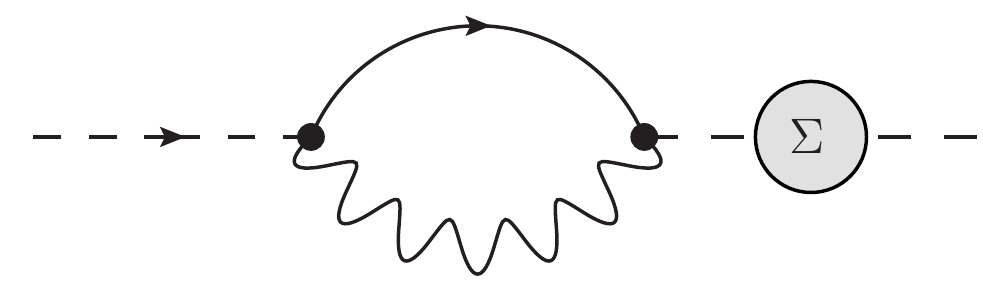},
\end{align}
which can be readily solved and we obtain Eq.~\eqref{eq:pathint:spcase:Gd} with the self-energy
\begin{equation}
 _{1}\Sigma(\omega) = \frac{U^2}{L} \sum_{k} {_{1}G^{0}_{\mathrm{w}}}(k;\omega) = \showgraph{width=1.5cm}{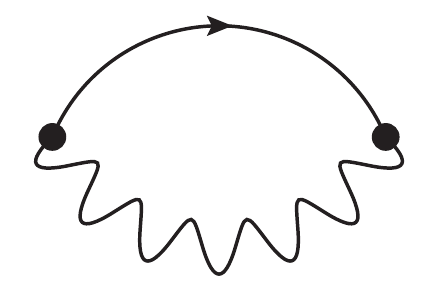}.
\end{equation}

By the same token, the full waveguide-Green's function is given by Eq.~\eqref{eq:pathint:spcase:Gwfull_w} 
and we can represent it in a diagrammatic form
\begin{equation}
 \label{dia:wGF-Dyson}
 {_{1}G_{\mathrm{w}}(k_f,k_i;\omega)} = \showgraph{width=1cm}{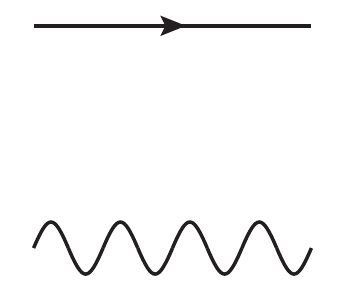} 
                               + \showgraph{width=3.5cm}{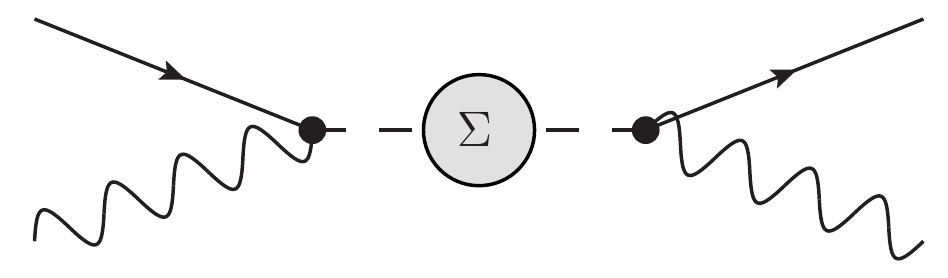}.
\end{equation}
This expression comprises two terms. The first term on r.h.s. corresponds to the free propagation 
of a photon and the TLS in the ground state while the second term on the r.h.s. describes the 
scattering off the (renormalized) TLS. Upon reinserting Eq.~\eqref{eq:FeynmanDiagrams:spcase:Gd_pertubationseries} 
into Eq.~\eqref{dia:wGF-Dyson}, we obtain the perturbation series of the full waveguide-Green's 
function. At this point, we would like to note that the full waveguide-Green's function still 
carries only one frequency, which is a manifestation of the fact that both, the photon and the 
ground state propagation, start and end at the same time.

\subsection{Two excitation sector}
\label{sec:FeynmanDiagrams:2pcase}
We now turn to the case where we have an additional photon in the system, i.e., we want to develop
the diagrammatic description of Sec.~\ref{sec:pathint:2pcase}. From Eq.~\eqref{eq:pathint:2pcase:Gd_pertubationseries} we know that the perturbation series for the TLS-Green's function consists of two Green's functions, which we will depict diagrammatically as

\begin{equation}\label{eq:FeynmanDiagrams:2pcase:Gw0}
 {_{2}G^{0}_{\mathrm{w}}}(k,k';t_f - t_i) = \showgraph{width=2cm}{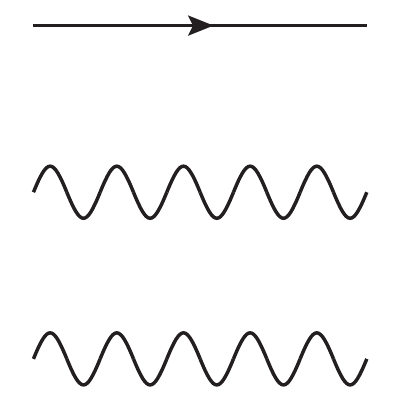}
\end{equation}
and
\begin{equation}\label{eq:FeynmanDiagrams:2pcase:Gdr}
 {_{2}G_{\mathrm{e},\mathrm{r}}}(k;t_f - t_i) = \showgraph{width=2cm}{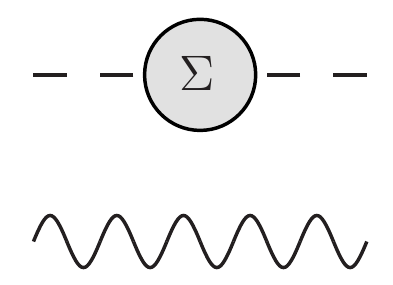}.
\end{equation}
Just as in the single-excitation case, these Green's functions describe the propagation of the 
excitations over a given time interval from $t_i$ to $t_f$ (this behavior can be immediately understood 
by the single frequency dependence in Eq.~\eqref{eq:pathint:2pcase:Gw0} and \eqref{eq:pathint:2pcase:Gdr}). 
The Green's function $_{2}G^{0}_{\mathrm{w}}(k,k';t_f - t_i)$ describes the simultaneous propagation of two 
photons with momenta $k$ and $k'$, together with a ground state field of the TLS. Similarly, the 
Green's function $_{2}G_{\mathrm{e},\mathrm{r}}(k;t_f - t_i)$ represents the propagation of an excited TLS and one
additional photon. In addition, in Eq.~\eqref{eq:FeynmanDiagrams:2pcase:Gdr} we have already 
encapsulated bubble-like renormalizations into the excited TLS propagation (as discussed 
in the single-excitation case).

With the help of these basic Green's function, we can now rewrite the TLS-Green's function 
in the two-excitation case as given by Eq.~\eqref{eq:pathint:2pcase:Gd_pertubationseries} within
the Feynman diagrammatic formulation as
\begin{align}\label{eq:FeynmanDiagrams:2pcase:Gd_pert}
 _{2}G_{\mathrm{e}}&(k,k',t_f - t_i) \nonumber \\
                              =-i \biggl\{& \showgraph{height=0.9cm}{2Gfdr.pdf} \nonumber \\
                              &+ \showgraph{height=1cm}{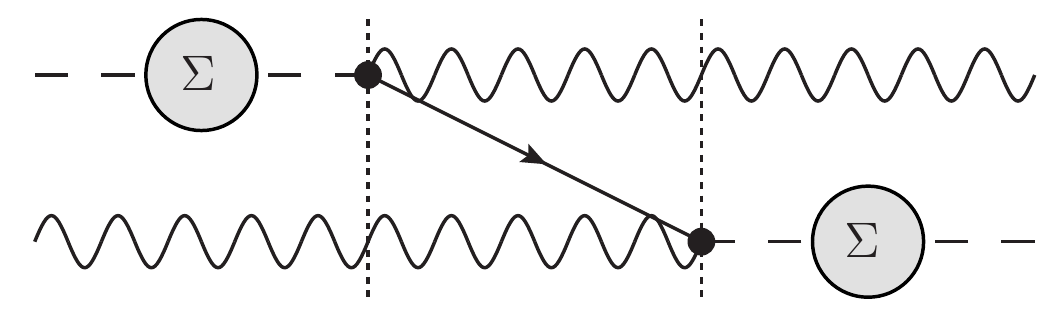} \nonumber \\
                              &+ \showgraph{height=1cm}{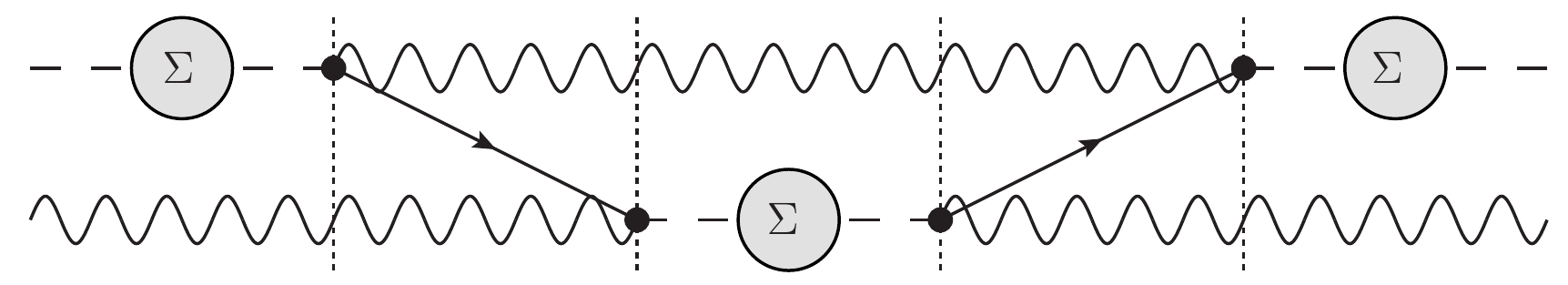} \nonumber \\
                              &+\dots \biggr\}.
\end{align}
Here, the dotted lines serves as indicator that separate distinct Green's functions from
each other and - in the time-domain formulation - imply an integration over the associated
intermediate times. Furthermore, the momentum of the free photon is conserved (as long as 
it does not interact with a vertex), each interaction vertex provides a factor $iU/\sqrt{L}$ 
and photons sandwiched between two interaction vertices imply a summation over the corresponding 
momentum (this applies for the upper, sandwiched photon in the third term, for example). 
For the interested reader, we have layed down the way to the equal-time Green's functions 
in Appendix \ref{sec:appendix:EqualTimeGF}.

This diagrammatic formulation of the TLS-Green's function's perturbation series in the 
two-excitation case provides a clear physical interpretation. The first term of the perturbation 
series corresponds to the situation when the TLS is excited and the additional photon is and 
stays free throughout the entire propagation. In the second term, the excited TLS emits a 
photon at some intermediate time, so for a given period of time two photons exist in the waveguide 
and propagate freely along with a free propagation of the TLS in the ground state. After a
certain time the initially free photon is absorbed by the TLS, whereas the other photon 
continues to propagate freely. Clearly, this induces correlations between the photons. In the 
higher-order terms of the perturbation series this process is repeated many times, which
effectively leads to photons that are emitted and reabsorbed, while the respective other 
photon is scattered by the TLS. Finally, we would like to note that all processes where the 
TLS directly reabsorbs the originally emitted photon without intermediate scattering are 
contained in the self-energy bubbles. In the case of a frequency-domain description, the 
intermediate convolutions integrals translate into multiplications and the common time-dependencies 
of the individual free-particle propagators that comprise the TLS-Green's function lead to a 
single frequency argument, just as has been the case for the single-excitation case.

The full waveguide-Green's function is given by Eq.~\eqref{eq:pathint:2pcase:Gw_result} and can 
be obtained by adding free waveguide-Green's functions to the full TLS-Green's function 
${_{2}G_{\mathrm{e}}(k,k',\omega)}$ in a symmetrized way (see Eq.~\eqref{eq:pathint:2pcase:Gw_result})
and by augmenting the perturbation series by a further term that takes into account the free 
propagation as described by Eq.~\eqref{eq:FeynmanDiagrams:2pcase:Gw0}. 

A general and very important property of the system is that the TLS cannot interact with free 
photons if it is in the excited state. It is exactly this property which renders the system
nonlinear, so that it is interesting to see how this feature translates into the diagrammatic 
formalism. For a linear system (e.g., a one-dimensional waveguide with a site-coupled bosonic 
quantum dot instead of the TLS) with two excitations, we could just take the square of the single 
excitation propagator, which, in our model, would lead to double excitations of the bosonic 
quantum dot. As already noted, the two-excitation Green's functions which we use have only one 
time dependence, so that the particles (photons (bosons) and the two fermions corresponding to 
the TLS in the ground and excited state, respectively) are created and annihilated at the same 
times. This means that the second term on the r.h.s. of Eq.~\eqref{eq:FeynmanDiagrams:2pcase:Gd_pert} 
can be interpreted as follows. A photon and an excited TLS are created at time $t_i$ and propagate 
up to an intermediate time $\tau$. At this time, the TLS emits an additional photon and goes to 
the ground state, until at time $\tilde{\tau}>\tau$ one photon is absorbed and the other one 
remains free. As all Green's functions in this series are retarded, at no point in time can 
double-excitation of the TLS take place. As a result, the few-photon nonlinearity emerges
and induces complex correlations between the photons 
\cite{Moeferdt-2013}.

\section{Properties of the Green's functions in the single-excitation sector}
\label{sec:spcase}
After having established the diagrammatic formulation of the theory in the single- and 
double-excitation case, we now turn to the examination of the single-excitation Green's functions
and will establish the connection of our framework to the other approaches discussed in Sec. 
\ref{sec:introduction}.

In Sec.~\ref{sec:model}, we have introduced the waveguide as a one-dimensional chain of length 
$L$ with nearest-neighbor hopping, thus exhibiting a cosine-shaped dispersion relation 
$\epsilon(k) = -2t \cos(k)$. Although we employ the cosine-shaped dispersion relation in this 
section, we may also use other dispersion relations (see the discussion in Sec.~\ref{sec:model}). 
Specifically, we will also consider a linear dispersion relation with group velocity $v$, i.e., 
$\epsilon_{\mu}(k) = \mu v k$, which is a good approximation for photons in the center of the 
cosine band. As already indicated in the dispersion relation, we then have to introduce a new 
quantum number, the chirality $\mu=R/L=+/-$ in order to account for the fact that we have both 
right- and left-moving photons. Furthermore, we pass from a set of discrete sites to a continuum 
description, which means that we replace all sums over real- or reciprocal space by corresponding 
integrals and replace $L$ by $2\pi$.

In the continuum limit, Eq.~\eqref{eq:pathint:spcase:Sigma} is given by
\begin{align}\label{eq:spcase:Sigma_general}
 {_{1}\Sigma(\omega)} = & U^2 \int \frac{\mathrm{d}k}{2\pi}  \frac{1}{\omega +\Omega/2 - \epsilon(k) + i 0} \nonumber \\
                      = & U^2 \mathcal{P} \int \frac{\mathrm{d}k}{2\pi} \frac{1}{\omega +\Omega/2 - \epsilon (k)} \nonumber \\
                        & - i \pi U^2 \int \frac{\mathrm{d}k}{2\pi} \sum_{n} \delta(k-k_n) 
                          \frac{1}{\partial \epsilon / \partial k} \biggr|_{k=k_n},
\end{align}
where $\mathcal{P}$ denotes the Cauchy principal value and the $k_n$ are given by the roots of 
$\omega + \Omega/2 = \epsilon(k)$. If all these roots are real (i.e. the energy is in the band), 
the principal value in Eq.~\eqref{eq:spcase:Sigma_general} becomes zero. Moreover, the second 
term can be identified with the density-of-states of the free waveguide $\nu(\omega)$. 
This gives
\begin{equation}\label{eq:spcase:Sigma_DOS}
  {_{1}\Sigma(\omega)} = -i \pi U^2 \nu(\omega).
\end{equation}
As a result, we find that the self-energy for the cosine dispersion as
\begin{equation}
 {_{1}\Sigma_{\text{cos}}(\omega,\Omega)} = \frac{U^2}{\omega +\Omega/2 + i \delta -2t} \sqrt{\frac{\omega +\Omega/2 + i \delta -2t}{\omega +\Omega/2 + i \delta +2t}},
\end{equation}
and for the linear dispersion as
\begin{equation}
 {_{1}\Sigma_{\text{lin}}(\omega)} = -i \frac{U^2}{v}.
\end{equation}
In the case of linear dispersion, the frequency-independent density-of-states leads to a 
frequency-independent self-energy. The self-energy of the cosine band exhibits a more complicated 
structure, it is purely imaginary when $\omega$ lies inside of the band and purely real when 
$\omega$ is outside the band. Note that the given representation of the self-energy is chosen in 
such a way that the square-root is evaluated at the correct side of the branch cut.
\begin{figure}[b]
 \begin{center}
  \includegraphics[width=0.45\textwidth]{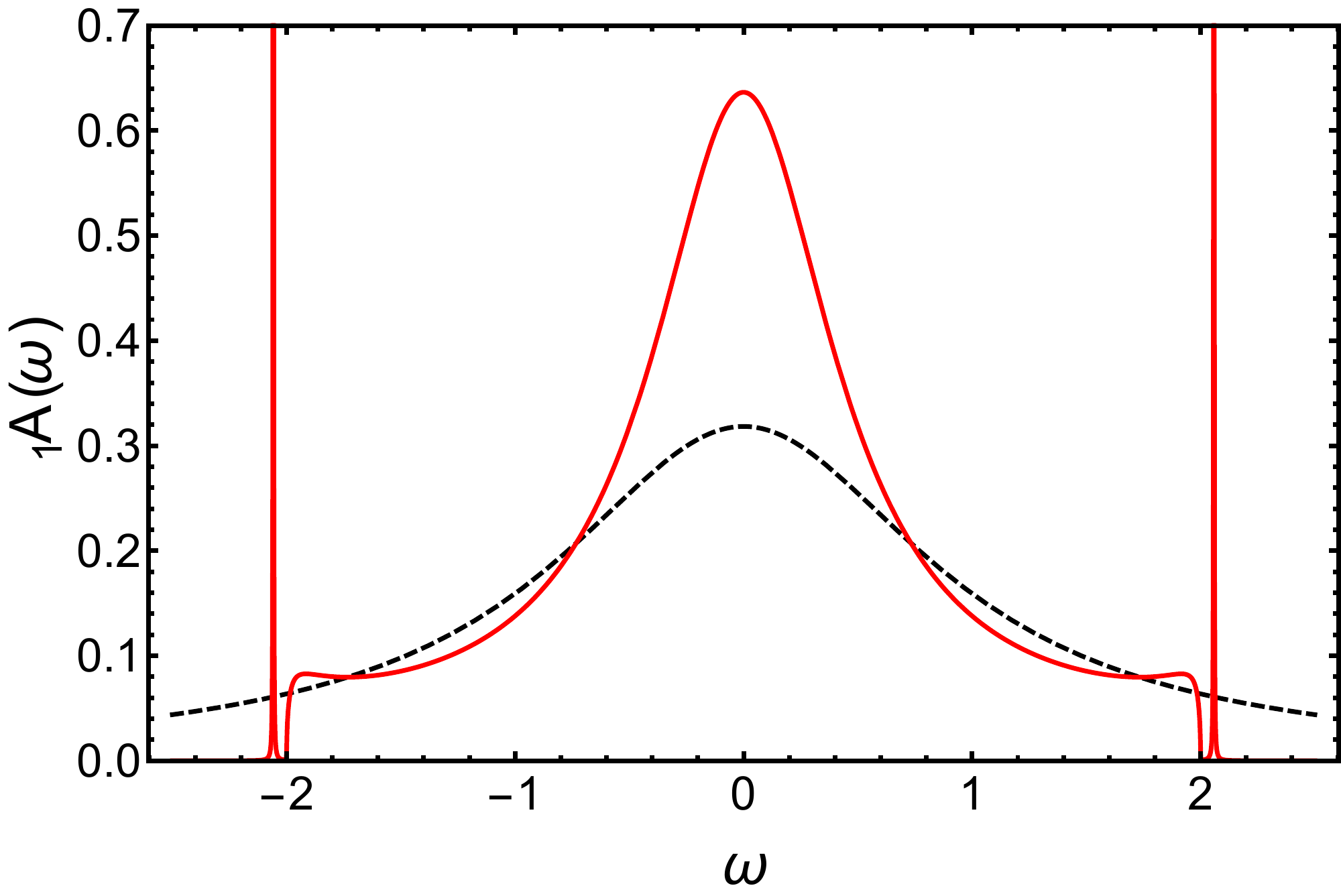}
 \end{center}
 \caption{Spectral density of the TLS with $\Omega=0$, $U=1$, $v=1$ and $t=1$ for the linear (black, dashed) and 
          cosine (red) dispersion relation. The spectral density of the cosine band clearly 
          displays the band edges and features spectrally ultra-sharp bound states in the 
          band gaps on either side of the band (when plotting the spectral density for the 
          cosine band, we have introduced an artificial broadening  $\delta=10^{-4}$ in 
          order to enhance the visibility of the bound states). By contrast, the spectral 
          density for the linear dispersion relation corresponds to a simple Lorentzian. }
 \label{fig:spcase:SpectralDensity}
\end{figure}

With these two self-energies, we are able to compute the spectral density of the TLS in
the single-excitation case as
\begin{equation}
 _{1}A(\omega) = -\frac{1}{\pi} \text{Im} \left[ {_{1}G_{\mathrm{e}}(\omega)} \right].
\end{equation}
In Fig.~\ref{fig:spcase:SpectralDensity}, we depict the spectral density for both dispersion 
relations. The spectral density for the linear dispersion relation is a simple Lorentzian with 
width $U^2/v$. In the case of the cosine dispersion, we clearly observe the frequency span of 
the band. In addition, we observe two sharp spectral lines in the band gaps on either side of
the band that correspond to two atom-photon bound states
\cite{ShiSun-2009,LongoSchmitteckertBusch-2010,LongoSchmitteckertBusch-2011}.

Furthermore, knowledge of the Green's functions enables us to compute the scattering matrix
($S$-matrix)
\begin{equation}\label{eq:spcase:SMatrix_def}
 S_{k,p} = \delta_{k,p} + i 2 \pi  \delta\left( \epsilon(k) - \epsilon(p) \right) T_{k,p},
\end{equation}
where we have obtained the transition matrix ($T$-matrix) via the Lehmann-Symanzik-Zimmermann 
(LSZ) reduction formula
\cite{LehmannSymanzikZimmermann-1955,ZarandBordaVonDelftAndrei-2004,BordaFritzAndreiZarand_2007},
\begin{equation}
 iT_{k,p} = -i G^{-1}_{0} (k) G(k,p) G^{-1}_{0}(p)\bigr|_{os}.
\end{equation}
In this expression, $G_{0}(k)$ and $G(k,p)$ denote, respectively, the free and the full, 
time-ordered Green's function. Further, the subscript $os$ indicates that the expression 
is taken on-shell, i.e., the scattering is elastic (or alternatively $\omega=\sum_i \epsilon(k_i)=\sum_f \epsilon(k_f)$, 
where the sums are over the initial and final momenta, respectively).
Using Eqs. \eqref{eq:pathint:spcase:Gw0_w} and \eqref{eq:pathint:spcase:Gwfull_w} for
the free and for the full Green's function, respectively, and omitting the free 
propagating part yields
\begin{equation}\label{eq:spcase:T_Gdot}
 iT_{k,p} = -i \frac{U^2}{2\pi} {_{1}G_{\mathrm{e}}}(\omega) \bigr|_{os}.
\end{equation}
We now rewrite the energy-conserving $\delta$-function that implements elastic scattering 
in terms of $\delta$-functions with momentum arguments and the density-of-states of the
free waveguide
\begin{equation}\label{eq:spcase:deltaE_deltak}
 \delta\left( \epsilon(k) - \epsilon(p) \right) = \pi \nu \left( \delta_{k,p} + \delta_{k,-p} \right),
\end{equation}
Upon combining Eqs. \eqref{eq:spcase:SMatrix_def} and \eqref{eq:spcase:T_Gdot} with
Eq.~\eqref{eq:spcase:deltaE_deltak} yields
\begin{equation}
 S_{k,p} = (1+r_k) \delta_{k,p} + r_k \delta_{k,-p},
\end{equation}
where the reflection amplitude $r_k$ is given by
\begin{equation}\label{eq:spcase:rk}
 r_k = \frac{-i\pi \nu U^2}{\epsilon(k) - \Omega/2 + i\pi \nu U^2}.
\end{equation}
In order to compare our results with the results from earlier works
\cite{ShenFan-2007-1,Nori-2008,ShiSun-2009}, 
we perform a shift of the energy $\omega \rightarrow \omega - \Omega/2$. Explicitly, for 
the linear dispersion relation we obtain 
\begin{equation}\label{eq:spcase:rk_lin}
 r^{\text{lin}}_k = \frac{-i U^2 / v}{vk - \Omega + i U^2 /v},
\end{equation}
while we obtain for the cosine dispersion relation
\begin{equation}
 r^{\text{cos}}_k = \frac{-i U^2}{2 t |\sin(k)|} \frac{1}{-2 t \cos(k) - \Omega + i U^2 / 2t |\sin(k)|}.
\end{equation}
Indeed, these expressions are in agreement with the results obtained in earlier
works by way of Bethe-Ansatz and LSZ-techniques 
\cite{ShenFan-2007-1,Nori-2008,ShiSun-2009}.

\section{Properties of the Green's functions in the two-excitation sector}
\label{sec:2pcase}
In correspondence with the single-excitation case, we now proceed to analyze the Green's functions
in the two-excitation case for the cosine and linear dispersion relations. In addition, we discuss 
the effect of band edges and bound photon-atom states on the perturbation series.

\subsection{Cosine dispersion relation - Discrete waveguide}
\label{sec:2pcase:Cosine}
The full Green's function of the TLS is given by Eq.~\eqref{eq:pathint:2pcase:Gd_pertubationseries} 
in the form of a perturbation series. This perturbation series can be cast in the form of a $T$-Matrix 
equation, given by Eq.~\eqref{eq:pathint:2pcase:Gd_Tmatrix}. For a discrete waveguide, we can solve
this equations simply by (numerical) matrix inversion and we defer the discussion of the continuum 
limit of the cosine band to sections \ref{sec:2pcase:Linear} (band center, approximately linear 
dispersion) and \ref{sec:2pcase:nonlinear} (band edge, approximately quadratic dispersion). 

\begin{figure}[t]
 \begin{center}
  \includegraphics[width=0.45\textwidth]{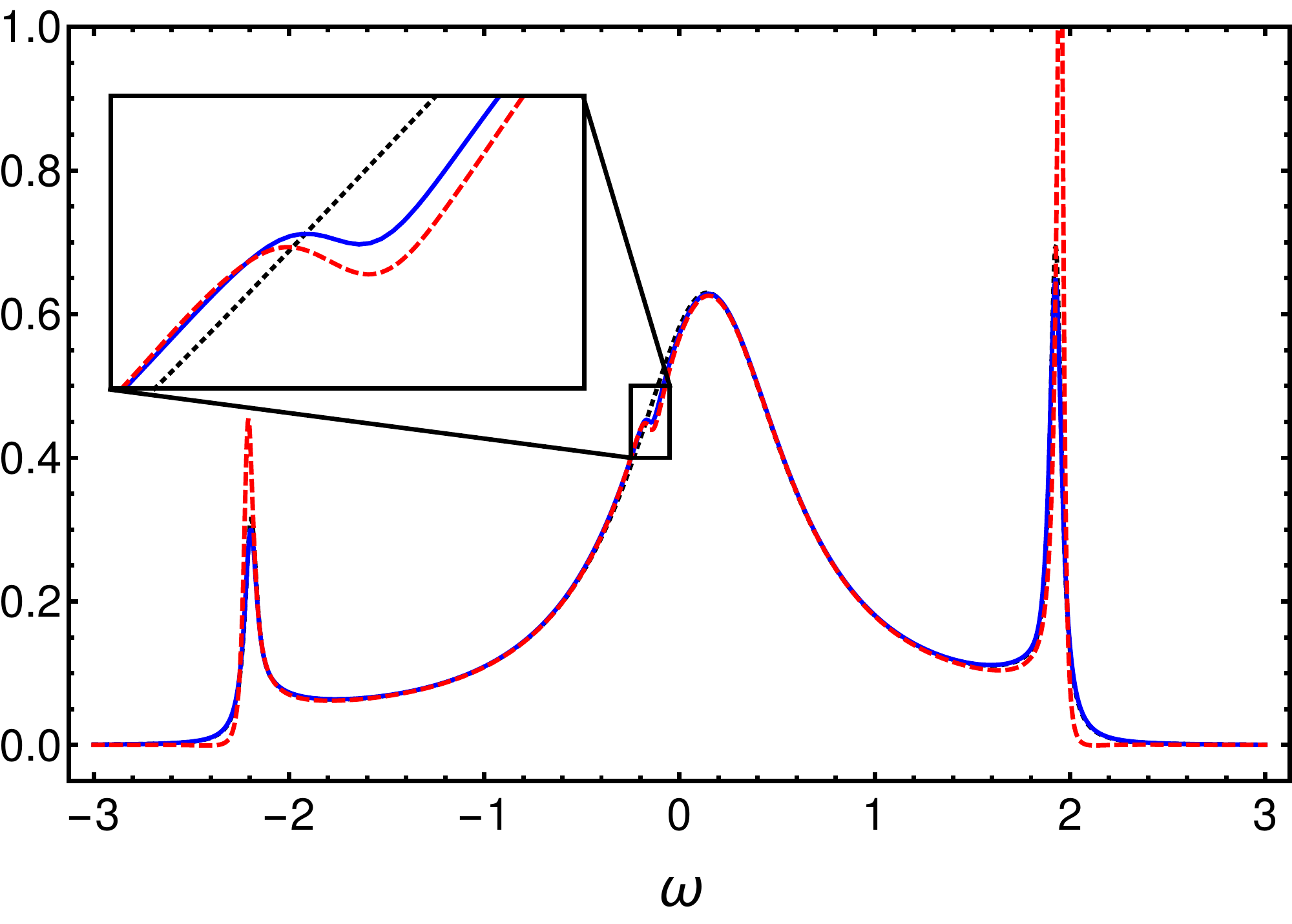}
 \end{center}
 \caption{Single and two-excitation spectral density $_{1}A(\omega)$ (black dotted) and $_{2}A(\pi/2,\omega)$ 
          (obtained via the Green's function approach (solid blue) and DMRG (dashed red)) of the TLS  with 
          $\Omega=0.3$ and $U=1$ for a tight-binding waveguide with $L = 600$ discrete sites and cosine 
          dispersion relation $\epsilon(k)=-2t\cos(k)$ ($t=1$). 
          In $_{2}A(\pi/2,\omega)$, we clearly observe a Fano-resonance just below $\omega = 0$
          (see text for details).
          This Fano-resonance is absent in the single-excitation spectral density of the TLS.
          When plotting the spectral densities, we have introduced an artificial broadening  
          $\delta=0.04$ in order to enhance the visibility of the Fano-resonance and to improve
          numerical convergence.
          }
 \label{fig:2pcase:Cosine:A2p}
\end{figure}

We define the two-excitation spectral density of the TLS as
\begin{equation}\label{eq:2pcase:Cosine:SpectralDensity}
 _{2} A (k,\omega) = -\frac{1}{\pi} \text{Im} \left[ {_{2}G_{\mathrm{e}}(k,k;\omega)} \right].
\end{equation}
In Fig.~\ref{fig:2pcase:Cosine:A2p}, we depict the two-excitation spectral density of the TLS, 
$_{2} A (\pi/2,\omega)$, and compare with the corresponding single-excitations spectral density, 
$_{1}A(\omega)$ (see Fig.~\ref{fig:2pcase:Cosine:A2p} for details of the systems). While both
spectral densities exhibit an overall similar behavior, we observe an additional feature in the
two-excitation spectral density which we attribute to a Fano resonance between the occupied
(renormalized) TLS and the additional photon in the waveguide. A Fano resonance appears when a broad 
continuum of states interacts with a single sharp mode. In our case, the (smeared out) TLS plays the 
role of the continuum whereas the additional photon acts as a sharp resonance.
We would like to note that we have introduced an artificial broadening of $\delta = 0.04$ in 
order to enhance the visibility of the Fano resonance and to improve numerical convergence of 
the matrix inversion. As a result, the bound states are not as sharp as in Fig.~\ref{fig:spcase:SpectralDensity} 
and the band edges are almost completely smeared out. Therefore, in order to make certain that 
the broadening does not introduce artificial features, we have confirmed the the results of 
the matrix inversion displayed in Fig.~\ref{fig:2pcase:Cosine:A2p} via computations of the 
spectral density by using an expansion in Chebyshev polynomials within the framework of the 
Density-Matrix-Renormalization-Group (DMRG) technique
as described in \cite{Braun_Schmitteckert:PRB14}.

In addition, we would like to stress that a similar Fano-resonance occurs in (analytically
solvable) case of linear dispersion (see Sec.~\ref{sec:2pcase:Linear}) so that we conclude 
that the occurrence of such a Fano-resonance between the occupied (renormalized) TLS and the 
additional photon in the waveguide is a generic feature of the few-photon nonlinearity in 
the $n$-photon-transport through a waveguide with embedded TLS for $n>1$.

Although we have solved Eq.~\eqref{eq:pathint:2pcase:Gd_Tmatrix} for a cosine dispersion relation
only, we would like to stress that our formalism is certainly not limited to this case (see the
discussion in section \ref{sec:model}). In fact, the $T$-matrix equation \eqref{eq:pathint:2pcase:Gd_Tmatrix}
is applicable to every possible dispersion relation that can be realized in a one-dimensional 
waveguide.

\subsection{Linear dispersion relation}
\label{sec:2pcase:Linear}
We now turn to our attention to the case of an (infinitely extended) linear dispersion 
relation $\epsilon_{\mu}(k) = \mu v k$ and thus ignore the effects of band edges, notably bound
photon-atom states and slow light regimes. Just as in the single-excitation case, we transit to
the continuum limit. Then, the self-energy of the renormalized TLS-Green's functions is
\begin{equation}
 {_{2}\Sigma}_{\text{lin}}(k;\omega) = -i \frac{U^2}{v},
\end{equation}
i.e., the same expression as in the single-excitation sector. This suggests that the linear 
dispersion exhibits certain special features so that, in contrast to the numerical treatment of 
the $T$-matrix, we aim at directly summing up the perturbation series, Eq.~\eqref{eq:pathint:2pcase:Gd_pertubationseries}. 
Formally, we can rewrite this series as
\begin{equation}\label{eq:2pcase:Linear:PerturbationSeries}
 {_{2}G_{\mathrm{e}}}(k_f,k_i;\omega) = \sum_{i} {_{2}G^{(i)}_{\mathrm{e}}}(k_f,k_i;\omega).
\end{equation}
Clearly, the first two terms of the series can be written down immediately, as no integration 
over internal momenta is required.

The third term, however, is given by
\begin{align}
 _{2}G^{(3)}_{\mathrm{e}}&(k_f,k_i;\omega) \nonumber \\
                  &= \frac{U^4}{2\pi} \int \frac{\mathrm{d}k}{2\pi} \, {_{2}G_{\mathrm{e},\mathrm{r}}}(k_i;\omega) {_{2}G^{0}_{\mathrm{w}}}(k_i,k;\omega) \nonumber \\
                  &\hphantom{= \frac{U^4}{2\pi} \int \frac{\mathrm{d}k}{2\pi}} \times  {_{2}G_{\mathrm{e},\mathrm{r}}}(k;\omega) \, {_{2}G^{0}_{\mathrm{w}}}(k,k_f;\omega) \nonumber \\
                  &\hphantom{= \frac{U^4}{2\pi} \int \frac{\mathrm{d}k}{2\pi}} \times {_{2}G_{\mathrm{e},\mathrm{r}}}(k_f;\omega),
\end{align}
where the integral only extends over the three internal Green's functions. As all these Green's 
functions have single poles that are shifted into the upper half plane, it follows that the 
integral and hence the entire third term in the series vanishes. As a matter of fact, this argument 
can be applied to terms of order higher or equal to three and this means that, in the case of a 
linear dispersion, the full TLS-Green's function is completely determined by the first two terms
\begin{align}\label{eq:2pcase:Linear:Gd}
 _{2}G&_{\mathrm{e}}(k_f,k_i;\omega) \nonumber \\
            = & \, {_{2}G_{\mathrm{e},\mathrm{r}}}(k_i;\omega) \, \delta_{k_i,k_f}\nonumber \\
              & + \frac{U^2}{2\pi} \, {_{2}G_{\mathrm{e},\mathrm{r}}}(k_i;\omega) \, {_{2}G^{0}_{\mathrm{w}}}(k_f,k_i;\omega) \, {_{2}G_{\mathrm{e},\mathrm{r}}}(k_f;\omega).
\end{align}

We are now in a position to provide a physical explanation for this termination of the 
perturbation series for linear dispersion relations after the first two terms by inspecting 
the first vanishing Feynman diagram (i.e., the third diagram on the r.h.s. of 
Eq.~\eqref{eq:FeynmanDiagrams:2pcase:Gd_pert})
in the time domain
\begin{equation}\label{eq:2pcase:Linear:Gdpic}
 {_{2}G^{(3)}_{\mathrm{e}}}(k_f,k_i;\omega) = \showgraph{height=1cm}{2Gfd_pert_2.pdf}
\end{equation}
The particle of interest is the (intermediary) upper photon, which is emitted and reabsorbed by 
the TLS. During the time that the upper photon "lives", the initial photon propagates and eventually 
gets absorbed. After a while, the photon is re-emitted and propagates further for a certain time. 
The entire process occupies a certain time $\tau>0$. During this time, the upper (intermediate) 
photon moves a certain distance due to the fixed group velocity $v>0$ of the linear dispersion 
and the absence of back-scattering mechanisms. This means that the photon has moved away from 
the TLS and actually cannot be reabsorbed, hence the diagram vanishes. We would like to note 
that this is a special property of the linear dispersion relation and does not hold for general 
dispersions relations, notably not near band edges and/or waveguide cut-off frequencies, i.e., 
in the vicinity of slow-light regimes.

\begin{figure}[t]
 \begin{center}
  \includegraphics[width=0.45\textwidth]{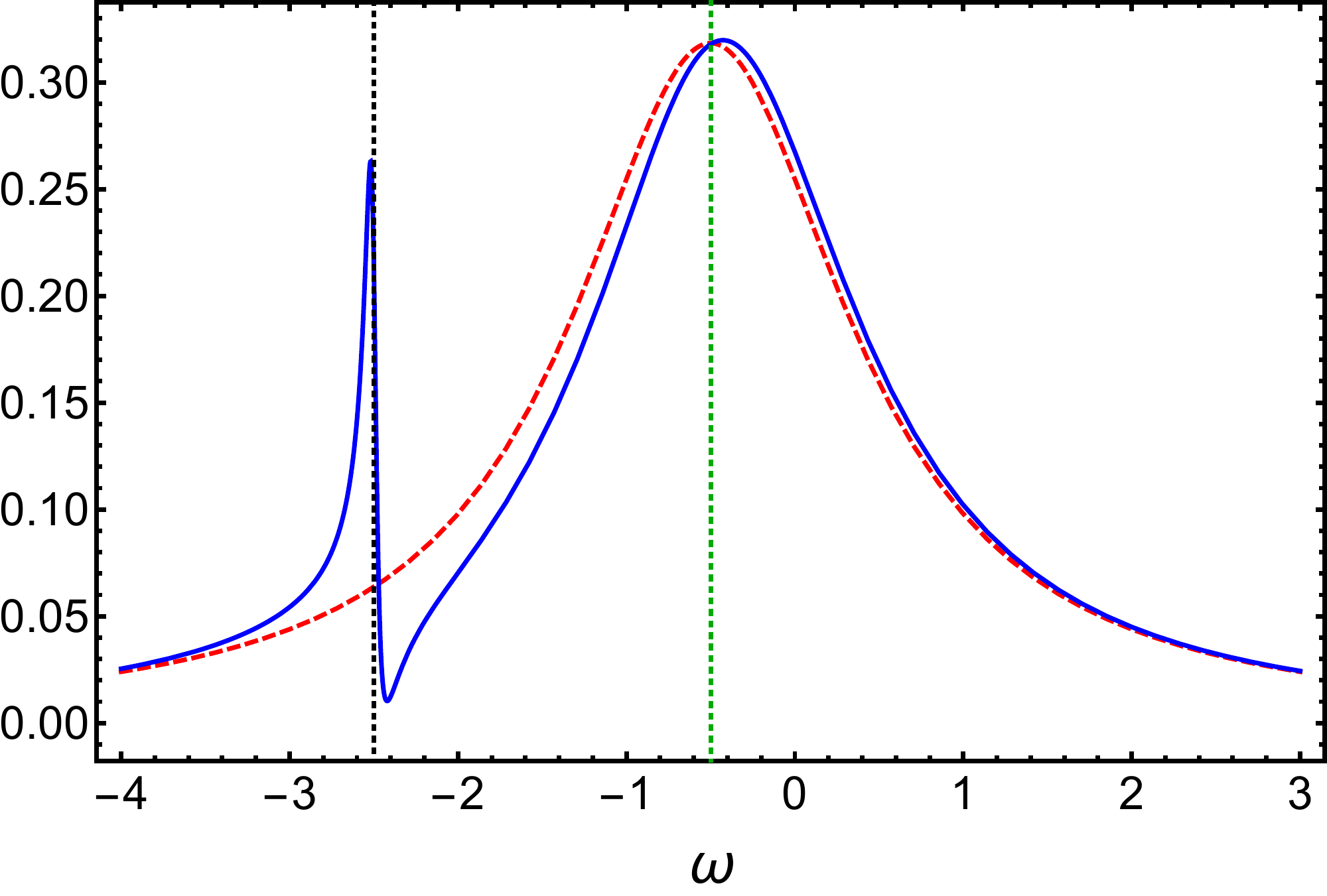}
 \end{center}
 \caption{Single- and two-excitation spectral densities $_{2}A(-1,\omega)$ (solid blue) and 
          $_{1}A(\omega)$ (dashed red) of the TLS with $\Omega=1$ and $U=1$ for a waveguide 
          with linear dispersion relation $\epsilon(k)=\mu vk$ and $v=1$ considered in the
          continuum limit. 
          We have shifted the single-excitation spectral density $_{1}A(\omega)$ by 
          $\omega \rightarrow \omega - vk$, so that the maxima of both plots overlap. The green 
          dotted line is at $vk + \Omega/2$ and indicates the maximum of $_{1}A(\omega)$.
          In $_{2}A(-1,\omega)$, we clearly observe a Fano-resonance at $\omega=2vk - \Omega/2$
          (black-dotted line) which is more pronounced than for the case of tight-binding
          waveguide in Fig.~\ref{fig:2pcase:Cosine:A2p}.
          As in the case of the tight-binding waveguide, this Fano-resonance is absent in the 
          single-excitation spectral density of the TLS.
          When plotting the spectral densities, we have introduced the same artificial broadening  
          $\delta=0.04$ as in the case of the tight-binding waveguide in order to enhance the 
          visibility of the Fano-resonance and make the graph comparable to that in
          Fig.~\ref{fig:2pcase:Cosine:A2p}.}
 \label{fig:2pcase:Linear:SpectralDensity}
\end{figure}
We have calculated the two-excitation spectral density according to Eq.~\eqref{eq:2pcase:Cosine:SpectralDensity} 
and depict the results together with the single-excitation spectral density in 
Fig.~\ref{fig:2pcase:Linear:SpectralDensity}. 
Similar to the numerical calculations for the cosine dispersion relation we, again, find again a 
Fano resonance located at $\omega_F = 2vk - \Omega/2$, which, however, is much more pronounced 
than in the cosine-shaped dispersion case. Comparing both dispersion relations, the Fano resonance 
appears in two different ways: In the case of the cosine dispersion relation the Fano resonance 
emerges as a result of the self-consistent treatment of the T-Matrix, whereas for the linear 
dispersion relation it can be traced back to the second term in the perturbation series 
$_{2}G^{(2)}_{\mathrm{e}}(k_f,k_i;\omega)$. To be more exact, the Fano resonance stems from the 
internal free waveguide Green's function ${_{2}G^{0}_{\mathrm{w}}}(k_i,k_f;\omega)$, which is of 
the form $(\omega - \omega_F + i\delta)^{-1}$. 
As a result, the Fano resonance is the consequence of a first order pole, regularized by a finite 
imaginary factor $i\delta$. For sufficiently small $\delta$, the spectral density can thus even
assume negative values. However, one has to take into account that we are not considering the 
spectral density of the full system, but only of a part of it (i.e., the part stemming from the 
TLS). Therefore, a negative spectral density is acceptable and can be considered as some sort of 
``gain'' (where the energy is taken from the waveguide), indicating effects such as  photon bunching
\cite{ShenFan-2007,ShenFan-2007-1,Moeferdt-2013}. 
Moreover, we would like to point out that the Fano resonance is less pronounced in Fig. 
\ref{fig:2pcase:Cosine:A2p}, although all energies (transition energy of the
TLS, photon energy) are in the linear regime of the cosine band. We attribute this regularization 
to the self-consistent treatment of the Green's function, which is not possible for the linear 
dispersion relation. 

In fact, the deeper reason behind the termination of the perturbation series for the linear dispersion 
is the separation of the photons on the Hamiltonian level into two kinds of photons (left- and 
right-moving ones). This changes the symmetry of the original Hamiltonian, i.e., the chirality
is introduced as a new quantum number, and eventually leads to the special analytic structure 
of the Green's functions and, as a consequence, to the termination of the perturbation series.

Finally, we can now construct the two-excitation $S$-matrix by generalizing the LSZ formalism 
presented in Sec.~\ref{sec:spcase}. Explicitly, the two-excitation $S$-Matrix is given as
\begin{align}\label{eq:2pcase:Linear:SMatrix_def}
  S_{k_{i}p_{i},k_{f}p_{f}} = & \left( \delta_{k_{i},k_{f}} \delta_{p_{i},p_{f}} + 
                                       \delta_{p_{i},k_{f}} \delta_{k_{i},p_{f}} \right) \nonumber \\
                              & + i 2\pi \delta(E) \,T_{k_{i}p_{i},k_{f}p_{f}},
\end{align}
where
\begin{equation}
 \delta(E) = \delta(\epsilon(k_{i}) + \epsilon(p_{i})  - \epsilon(k_{f})- \epsilon(p_{f})) 
\end{equation}
ensures elastic scattering and the associated $T$-matrix is defined as
\begin{align}
 i T&_{k_{i}p_{i},k_{f}p_{f}} \nonumber \\
    & = - i G^{-1}_{0}(k_{i},p_{i}) \, G(k_{i},p_{i};k_{f},p_{f}) \, G^{-1}_{0}(k_{f},p_{f}) \bigr|_{os}.
\end{align}
In this expression, $G_{0}(k,p)$ and $G(k,p;k',p')$ denote, respectively, the free and the full 
waveguide Green's function. With the help of Eq.~\eqref{eq:pathint:2pcase:Gw_result}, we explicitly find
\begin{equation}\label{eq:2pcase:Linear:TMatrix}
 i T_{k_{i}p_{i},k_{f}p_{f}} = -i \frac{U^2}{2\pi} {_{2}G^{\text{sym}}_{\mathrm{e}}}(k_i,p_i,k_f,p_f;\omega) \bigr|_{os},
\end{equation}
where ${_{2}G^{\text{sym}}_{\mathrm{e}}}(k_i,p_i,k_f,p_f;\omega)$ is defined by Eq.~\eqref{eq:pathint:2pcase:Gd_sym}. 

In order to compare these results with results of earlier works, we again perform a frequency shift
$\omega \rightarrow \omega - \Omega/2 $. We carry out the actual calculations of the $S$-Matrix in 
Appendix \ref{sec:appendix:2pSMatrix} so that we report here only the final results. For different 
chirality configurations, the $S$-matrix reads
\begin{itemize}
 \item $ k_{i}^{R} p_{i}^{R} \rightarrow k_{f}^{R} p_{f}^{R} $
 \begin{align}
  S^{RR,RR}_{k_{i}p_{i},k_{f}p_{f}} = &t_{k_i} t_{p_i} \left( \delta_{k_i,k_f} \delta_{p_i,p_f} + \delta_{k_i,p_f} \delta_{p_i,k_f} \right) \nonumber \\
                             & + S^{2,\text{P.V.}}_{k_{i}p_{i},k_{f}p_{f}},
 \end{align}
 \item $ k_{i}^{R} p_{i}^{R} \rightarrow k_{f}^{R} p_{f}^{L} $
 \begin{align}
  S^{RR,RL}_{k_{i}p_{i},k_{f}p_{f}} = &t_{k_i} r_{p_i} \delta_{k_i,k_f} \delta_{p_i,-p_f} + r_{k_i} t_{p_i} \delta_{k_i,-p_f} \delta_{p_i,k_f}  \nonumber \\
                             & + S^{2,\text{P.V.}}_{k_{i}p_{i},k_{f}p_{f}},
 \end{align}
 \item $ k_{i}^{R} p_{i}^{R} \rightarrow k_{f}^{L} p_{f}^{L} $
 \begin{align}
  S^{RR,LL}_{k_{i}p_{i},k_{f}p_{f}} = &r_{k_i} r_{p_i} \left( \delta_{k_i,-k_f} \delta_{p_i,-p_f} + \delta_{k_i,-p_f} \delta_{p_i,-k_f} \right) \nonumber \\
                             & + S^{2,\text{P.V.}}_{k_{i}p_{i},k_{f}p_{f}}.
 \end{align}
\end{itemize}
In these expressions, the superscript of the momenta indicates the chirality, $r_k$ is the single-excitation 
reflection amplitude (c.f. Eq.~\eqref{eq:spcase:rk_lin}), $t_k = 1 + r_k$ is the single-excitation transmission 
amplitude and $S^{2,\text{P.V.}}_{k_{i}p_{i},k_{f}p_{f}}$ is given by 
\begin{align}
 S^{2,\text{P.V.}}_{k_{i}p_{i},k_{f}p_{f}} = & \frac{iU^4}{\pi v}\delta_{k_i+p_i,k_f+p_f} \nonumber \\ 
                                             & \times \frac{\left( k_i + p_i - 2 \Omega + i U^2/v \right)}{(v p_i - \Omega +i U^2/v)(v k_i - \Omega +i U^2/v)} \nonumber \\
                                             & \times \frac{1}{(v p_f - \Omega +i U^2/v)(v k_f - \Omega +i U^2/v)}.
\end{align}
Our results are thus in accordance with the results obtained by other techniques in earlier works 
\cite{ShenFan-2007,ShenFan-2007-1,ShiSun-2009}. 

Within our scheme, we can also give a new explanation of the term $S^{2,\text{P.V.}}_{k_{i}p_{i},k_{f}p_{f}}$. 
In the first place, this term appears by replacing the free waveguide-Green's function in $_{2}G_{\mathrm{e}}^{(2)}$ 
by the Dirac identity
\begin{equation}
 \frac{1}{\omega - v k - v p +i 0} = \mathcal{P}\frac{1}{\omega - v k - v p} - i\pi \delta\left( \omega - v k - v p \right).
\end{equation}
The two terms in the Dirac identity can be interpreted as follows. The imaginary part that is proportional 
to a $\delta$-function corresponds to long-time, real processes, because the $\delta$-function sets the 
particles on-shell. The real part that contains the principal value, however, does not place a constraint 
on the momenta. The momenta can be chosen freely and are only restricted by energy conservation. Therefore,
this term corresponds to short times, which are on the scale of the Heisenberg uncertainty principle, i.e.,
they correspond to \emph{virtual processes}.

\subsection{Band edge effects}
\label{sec:2pcase:nonlinear}
Finally, we address the case of band edges by following the same line of reasoning as in section 
\ref{sec:2pcase:Linear}. Perhaps the simplest nonlinear dispersion relations exhibiting a band edge
is the quadratic dispersion relations $\epsilon(k)=tk^2$ with $t>0$. In this case, the self-energy is
given by
\begin{equation}
 {_{2}\Sigma}_{\text{qu}}(k;\omega) = -i \frac{\pi U^2}{\sqrt{t(\omega-t k^2 + i0)}},
\end{equation}
where we scaled out the factor $\Omega/2$ again. 
The first two terms of the perturbation series given by Eq.~\eqref{eq:2pcase:Linear:PerturbationSeries} 
can again be computed straightforwardly, but the third term (and all higher-order terms) exhibit different 
characteristics. First of all, the quadratic dispersion relation induces poles on both sides of the 
complex half-plane, which means that the integral over internal momenta is not vanishing, hence 
$_{2}G^{(3)}_{\mathrm{e}}(k_f,k_i;\omega)$ is finite. This was expected, since the quadratic dispersion relation 
exhibits a state where the group velocity $v_g=0$, which means that an emitted photon can be reabsorbed 
by the TLS after a certain amount of time (c.f. discussion in section \ref{sec:2pcase:Linear}). Secondly, 
the self-energy ${_{2}\Sigma}_{\text{qu}}(k;\omega)$ leads to two branch cuts, one in each half space of 
the complex plane. These branch cuts represent major obstacles in the integration over the internal momenta
and we have been unable to find a closed form-solution for $_{2}G^{(3)}_{\mathrm{e}}(k_f,k_i;\omega)$. 

From a more physical point of view, however, we expect that the higher-order processes encapsulate the 
effect of interaction-induced radiation trapping (IIRT)
\cite{LongoSchmitteckertBusch-2010,LongoSchmitteckertBusch-2011}. This phenomenon describes the 
excitation of the atom-photon bound state by a two-photon pulse as the result of a nonlinear process.
 This expectation can be motivated by the diagrammatic form of ${_{2}G}^{(3)}_{\mathrm{e}}(k_f,k_i;\omega)$,

\begin{center}
 \includegraphics[height=1cm]{2Gfd_pert_2.pdf}.
\end{center}
In the top line the TLS emits a photon, which is reabsorbed at a later time. This is exactly the 
behavior one would expect from the atom-photon bound state, since the radiation cannot leave the 
TLS.

\begin{figure}[t]
 \centering
 \includegraphics[width=.45\textwidth]{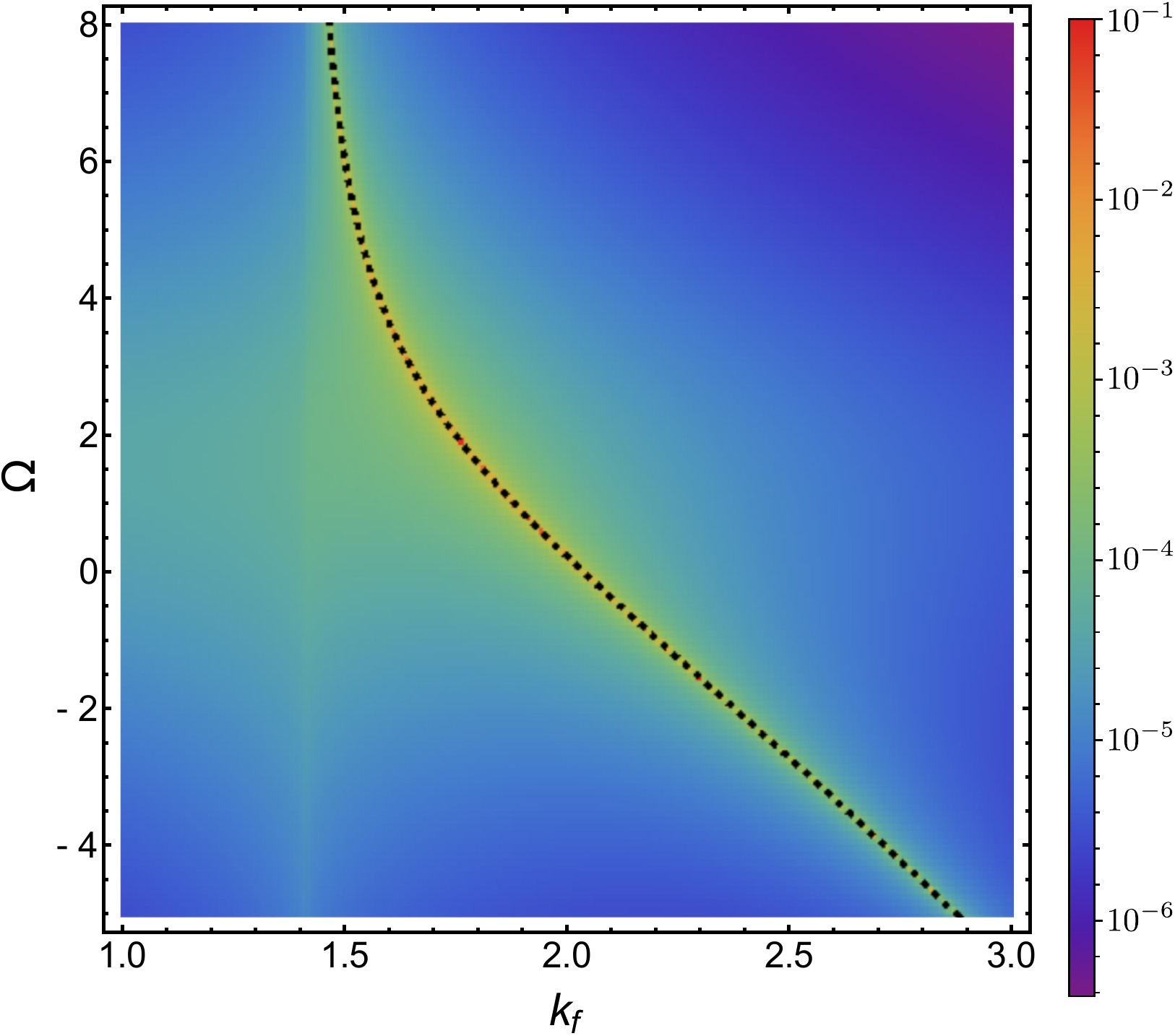}
 \caption{Logarithmic plot of $\left|{_{2}G}^{(3)}_{\mathrm{e}}(k_f,k_i;\omega)\right|$ with 
          $k_i=p_i=1$, $t=1$, $U=1$, $\omega = \epsilon(k_1) + \epsilon(k_2)$. We have added 
          a small imaginary part $\delta=10^{-3}$ to $\omega$ for an artificial broadening 
          of the resonance. The black dotted line represents the solution of 
          $\omega = \omega_{\mathrm{BS}} + \epsilon(k_f)$. The resonance approaches 
          $k_f \rightarrow \sqrt{2}$ for large values of $\Omega$, since the bound state energy 
          $\omega_{BS} \rightarrow 0$ in this case.}
 \label{fig:2pcase:BSPlot}
\end{figure}

\begin{figure}[t]
 \begin{center}
  \includegraphics[width=.45\textwidth]{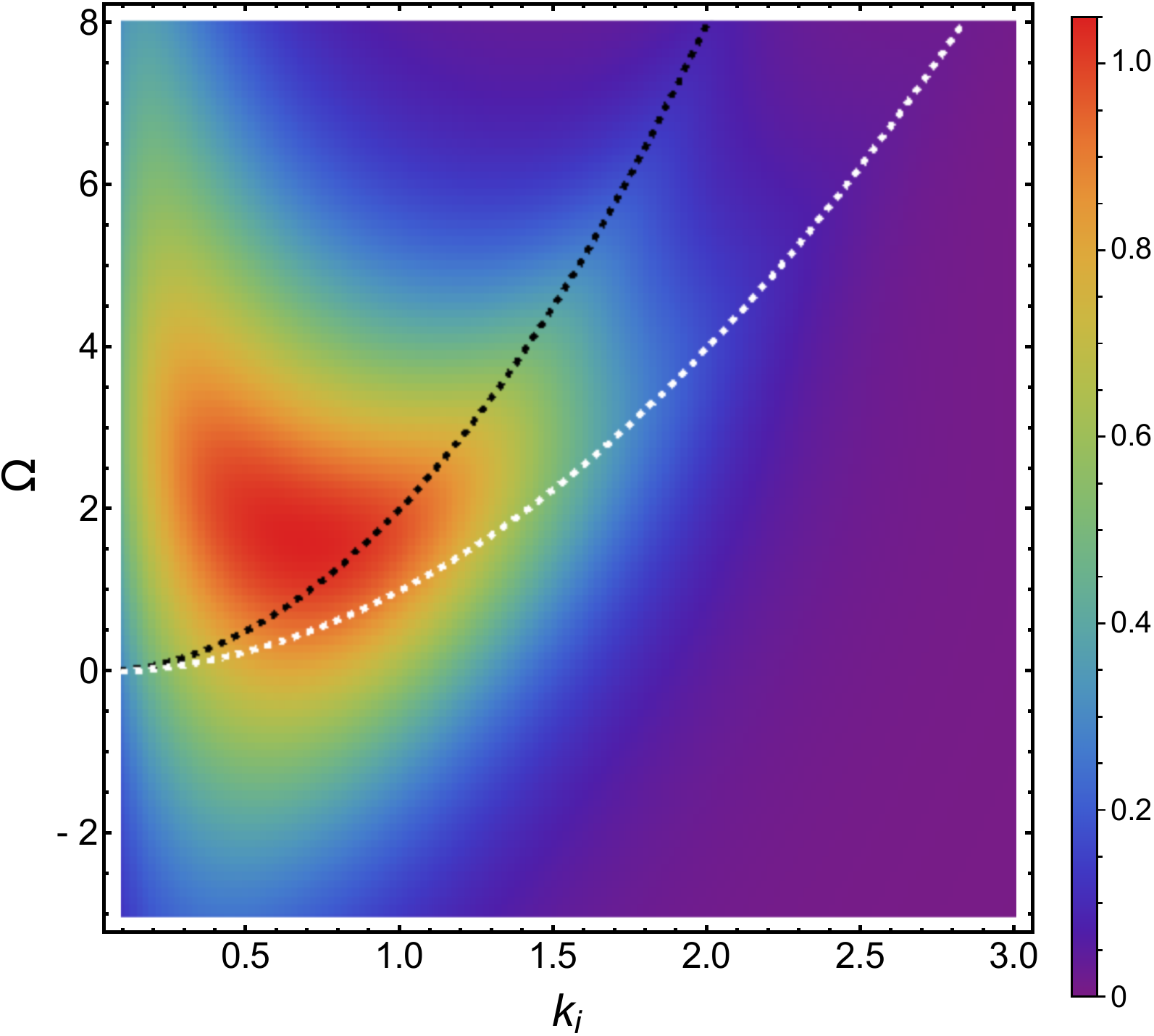}
 \end{center}
 \caption{Plot of $\left|\text{Res} \left[{_{2}G}^{(3)}_{\mathrm{e}}, k_0 \right]\right|$ for two 
          identical initial photons $\omega=2\epsilon(k_i)$, $U=1$, $t=1$, $\delta=10^{-4}$ and 
          $\eta=10^{-6}$. The black and white lines represent the parabolas $\Omega=2\epsilon(k_i)$ 
          and $\Omega=\epsilon(k_i)$, respectively.}
 \label{fig:2pcase:BSWeightPlot}
\end{figure}

The prototypical process of IIRT includes two initial photons, which are transformed into an 
atom-photon bound state and a photon with a different momentum. Energetically, this process 
is described by
\begin{equation}\label{eq:2pcase:Quad:BSCondition}
 \omega = \epsilon(k_i) + \epsilon(p_i) = \omega_{\mathrm{BS}} + \epsilon(k_f),
\end{equation}
where $k_i$ and $p_i$ are the momenta of the initial photons, $k_f$ is the momentum of the final 
photon, $\omega$ is the total energy and $\omega_{\mathrm{BS}}$ is the energy of the bound 
atom-photon state, which can be found by solving the equation
\begin{equation}
 \omega_{\mathrm{BS}} - \Omega + i \frac{U^2 \pi}{\sqrt{t \omega_{\mathrm{BS}}}} = 0.
\end{equation}
In fact, setting ${_{2}G}^{(3)}_{\mathrm{e}}(k_f,k_i;\omega)$ on-shell (thus making it proportional 
to a scattering matrix element) and computing the integration over the internal momenta numerically 
yields a sharp peak at precisely the momenta $k_f$ which fulfill 
$\omega = \omega_{\mathrm{BS}} + \epsilon(k_f)$ (cf. Fig.~\ref{fig:2pcase:BSPlot}). 
This indicates that ${_{2}G}^{(3)}_{\mathrm{e}}(k_f,k_i;\omega)$ (together with the higher-order 
processes) describes indeed the physics behind the IIRT. Moreover, treating the sharp peak as a 
pole (which is motivated by the fact that the width of the resonance scales with the small imaginary 
part $i\delta$) enables us to compute the residue of the resonance via
\begin{equation}
 \text{Res} \left[{_{2}G}^{(3)}_{\mathrm{e}}, k_0 \right] = \frac{1}{2\pi i} \oint_{C} \mathrm{d}z \,{_{2}G}^{(3)}_{\mathrm{e}}(z,k_i;\omega)\bigr|_{os},
\end{equation}
where $k_0$ is the solution of Eq.~\eqref{eq:2pcase:Quad:BSCondition}, the subscript $os$ indicates 
that the expression is taken on-shell and the contour $C$ is a circle centered at $k_0$ with radius 
$\eta$. We have plotted the residue in Fig.~\ref{fig:2pcase:BSWeightPlot}, together with the 
conditions that the two initial photons are on resonance with the TLS individually 
($\Omega = \epsilon(k_i)$) and collectively (\mbox{$\Omega = \omega = 2\epsilon(k_i)$}). We can see 
here that the strength of the pole is sensitive to the TLS being on resonance with the collective 
photonic excitation, rather than with the individual ones. This is another indicator that IIRT emerges 
as the result of the nonlinear behavior of the TLS in the presence of two or more photons. 
Furthermore, this result shows that an analytic solution for ${_{2}G}^{(3)}_{\mathrm{e}}(k_f,k_i;\omega)$ 
is still highly desirable as this would provide further insights into the physics of IIRT.

Finally, we would like to conclude this section by recalling the physical interpretation of each term 
in the perturbation series of ${_{2}G}_{\mathrm{e}}(k_f,k_i;\omega)$: The first term describes the 
single-photon scattering and does not induce correlations between the two photons. The second term 
gives rise to photon bunching and is always finite, independent of the dispersion relation. The third 
term and all higher-order terms are nontrivial, since they include integrations over internal momenta. 
In the case of a linear dispersion relation their contribution vanishes. Conversely, these higher-order
terms become particularly relevant for frequencies near band edges and/or waveguide cut-offs and lead
to the effect of IIRT.

\section{Conclusion}
\label{sec:Conclusion}

In summary, we have developed an efficient and flexible quantum field theoretical approach
to few-photon transport problems in one-dimensional photonic waveguides with embedded
quantum impurities. Our approach is based on a coherent-state path integral formulation 
which allows us to formulate a Feynman diagram representation that elucidates the nature
of the underlying physical processes. For instance, our framework allows for both frequency- 
and time-domain considerations and we have utilized both of them throughout this manuscript
in order to arrive at several novel physical explanations.

In the case of a single excitation, our approach delivers closed-form analytical expressions 
for the Green's functions for arbitrary dispersion relations and we have computed spectral 
densities and scattering matrices. Similarly, for the case of two excitations, our approach 
demonstrates that and why all diagrams of order higher than two vanish for systems with linear 
dispersion, again facilitating a closed-form expressions for the Green's function and all 
physical quantities. For arbitrary dispersion relations, the case of two excitations does 
not admit general closed-form solutions and we have derived a self-consistent $T$-matrix 
equation that reduces the required computational effort to that of a single-excitation 
calculation.

The results of our approach are consistent with results from several earlier works in which 
different techniques have been employed
\cite{ShenFan-2007,ShenFan-2007-1,ShiSun-2009}.
In addition, for the two-excitation case, we have derived a Fano-resonance effect in the 
spectral density that originates from an interference between the occupied (renormalized) 
TLS and the additional photon in the waveguide. This Fano resonance represents a unique 
signature of the few-photon nonlinearity inherent in our systems. Furthermore, we have
shown that while linear dispersion relations are in some sense limited to photon-correlation 
effects only, dispersion relations with slow light regimes (i.e., with band edges and/or 
waveguide cut-off frequencies) feature additional effects related to interaction-induced
radiation trapping effects that originate from nonlinear scattering processes that involve
bound atom-photon states. This leads to very rich physics even for such -- on first sight --
rather simple systems.

Finally, our framework can serve as the basis for further studies that involve more complex 
scenarios such as several and many-level quantum impurities, networks of coupled waveguides, 
disordered systems, and non-equilibrium effects. 

\section{Acknowledgments}
M.P.S., T.S., and K.B. gratefully acknowledge fruitful discussions with Prof. D. Ebert.

\appendix

\section{Green's functions in the single excitation sector}
\label{sec:appendix:spGF}
In this appendix, we complete the derivation of the single-excitation Green's function. 
The full TLS- and waveguide-Green's functions have been derived in Sec.~\ref{sec:pathint:spcase} 
and Sec.~\ref{sec:FeynmanDiagrams:spcase}, but the case where a photon is absorbed or emitted 
by the TLS is still missing. The general expression for the absorption-Green's function reads
\begin{align}\label{eq:appendix:spGF:Gda_bosons}
 {_{1}G_{\mathrm{ab}}(k_i;t_f-t_i)}  = &-i \int \prod_{k,k'} \mathrm{d}\phi_{f,k} \mathrm{d}\phi^*_{f,k} 
                                                              \mathrm{d}\phi_{i,k'}\mathrm{d}\phi^*_{i,k'} \nonumber \\        
	                                        & \times \phi^{*}_{i,k_i}e^{-\sum_{k} \phi_{f,k}\phi^*_{f,k} 
	                                                                    -\sum_{k} \phi_{i,k}\phi^*_{i,k}} \nonumber \\ 
                                          & \times G_{\mathrm{ab}}(f,i,t_f-t_i),
\end{align}
where $G_{\mathrm{ab}}(f,i,t_f-t_i)$ is given by Eqs. \eqref{eq:pathint:G_MatrixStructure} and 
\eqref{eq:pathint:Projection}. Performing the same integrations as in Sec.~\ref{sec:pathint} 
yields
\begin{equation}
 {_{1}G_{\mathrm{ab}}(k_i;\omega)} = \frac{U}{\sqrt{L}} {_{1}G^{0}_{\mathrm{w}}(k_i;\omega)} {_{1}G_{\mathrm{e}}(\omega)},
\end{equation}
where, ${_{1}G^{0}_{\mathrm{w}}(k_i;\omega)}$ and ${_{1}G_{\mathrm{e}}(\omega)}$ are defined in Eqs. 
\eqref{eq:pathint:spcase:Gw0_w} and \eqref{eq:pathint:spcase:Gd}, respectively. The diagrammatic 
representation of this process is straightforward. A free photon is annihilated and excites the 
TLS
\begin{equation}
 {_{1}G_{\mathrm{ab}}(k_i;\omega)} = \showgraph{height=1.2cm}{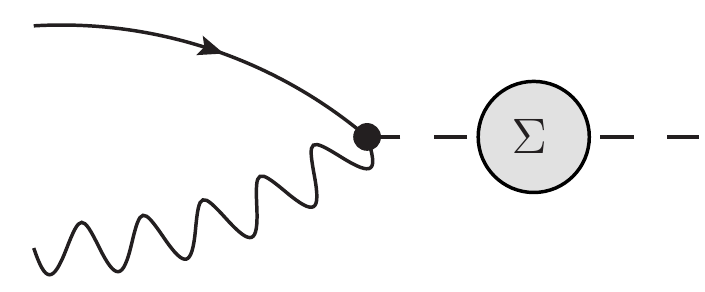}.
\end{equation}
The calculation of the emission-Green's function follows the exact same path, the starting 
expression being
\begin{align}\label{eq:appendix:spGF:Gde_bosons}
{_{1}G_{\mathrm{em}}(k_f;t_f-t_i)}  = &-i \int \prod_{k,k'} \mathrm{d}\phi_{f,k} \mathrm{d}\phi^*_{f,k} 
                                                              \mathrm{d}\phi_{i,k'}\mathrm{d}\phi^*_{i,k'} \nonumber \\        
	                                        & \times \phi_{f,k_f}e^{-\sum_{k} \phi_{f,k}\phi^*_{f,k} 
	                                                                -\sum_{k} \phi_{i,k}\phi^*_{i,k}} \nonumber \\ 
                                          & \times G_{\mathrm{em}}(f,i,t_f-t_i).
\end{align}
In turn, this yields
\begin{equation}
 {_{1}G_{\mathrm{em}}(k_f;\omega)} = \frac{U}{\sqrt{L}} {_{1}G_{\mathrm{e}}(\omega)}{_{1}G^{0}_{\mathrm{w}}(k_f;\omega)},
\end{equation}
together with a straightforward diagrammatic representation. An excited TLS emits a photon and 
finds itself in the ground state
\begin{equation}
 {_{1}G_{\mathrm{em}}(k_f;\omega)} = \showgraph{height=1.2cm}{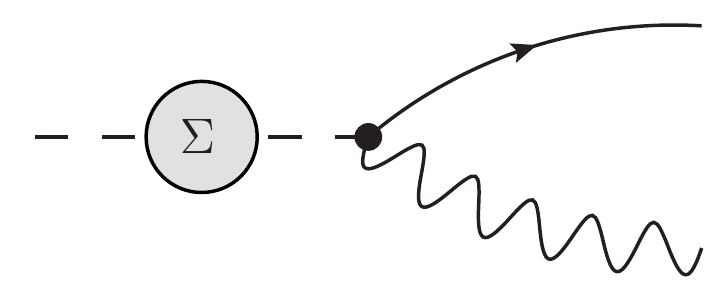}.
\end{equation}

\section{Green's functions in the two excitation sector}
\label{sec:appendix:2pGF}
Just as in the single-excitation case, we can also derive absorption- and emission-Green's 
functions for the case of two excitations. However, we will omit the representation of the 
Green's functions by Feynman diagrams, since they include certain symmetrizations such as
those in section \ref{sec:FeynmanDiagrams:2pcase}. 

Explicitly, the absorption Green's function is given by
\begin{align}\label{eq:appendix:2pGF:Gda_bosons}
{_{2}G_{\mathrm{ab}}}  & (k_i,p_i,k_f;t_f-t_i)  \nonumber \\
            = & -i \int \prod_{k,k'} \mathrm{d}\phi_{f,k} \mathrm{d}\phi^*_{f,k} 
                                  \mathrm{d}\phi_{i,k'}\mathrm{d}\phi^*_{i,k'} \nonumber \\        
	            & \times \phi^{*}_{i,k_i} \phi^{*}_{i,p_i} \phi_{f,k_f} e^{-\sum_{k} \phi_{f,k}\phi^*_{f,k} 
	                                                                       -\sum_{k} \phi_{i,k}\phi^*_{i,k}} \nonumber \\ 
              & \times G_{\mathrm{ab}}(f,i,t_f-t_i),
\end{align}
where $G_{\mathrm{ab}}(f,i,t_f-t_i)$ is given by Eqs. \eqref{eq:pathint:G_MatrixStructure} and \eqref{eq:pathint:Projection}. 

Integrating out the bosonic fields results in
\begin{align}
 {_{2}G_{\mathrm{ab}}}&(k_i,p_i,k_f;\omega) \nonumber \\
      = & \frac{U}{\sqrt{L}} \, {_{2}G^0_{\mathrm{w}}}(k_i,p_i;\omega) \nonumber \\
        & \times \left[ {_{2}G_{\mathrm{e}}(k_f,k_i;\omega)} + {_{2}G_{\mathrm{e}}(k_f,p_i;\omega)} \right].
\end{align}
The physical process is analogous to the single-excitation case. Two photons are initialized 
in the waveguide and one of them is absorbed by the TLS. 

By the same token, the general formula for the emission-Green's function reads
\begin{align}\label{eq:appendix:2pGF:Gde_bosons}
{_{2}G_{\mathrm{em}}}  &  (k_i,k_f,p_f;t_f-t_i)  \nonumber \\
                = & -i\int \prod_{k,k'} \mathrm{d}\phi_{f,k} \mathrm{d}\phi^*_{f,k} 
                                      \mathrm{d}\phi_{i,k'}\mathrm{d}\phi^*_{i,k'} \nonumber \\        
	                & \times \phi^{*}_{i,k_i} \phi_{f,k_f} \phi_{f,p_f} e^{-\sum_{k} \phi_{f,k}\phi^*_{f,k} 
	                                                                       -\sum_{k} \phi_{i,k}\phi^*_{i,k}} \nonumber \\ 
                  & \times G_{\mathrm{em}}(f,i,t_f-t_i).
\end{align}
Upon integrating out the bosonic fields, we finally find
\begin{align}
 {_{2}G_{\mathrm{em}}}&(k_i,k_f,p_f;\omega) \nonumber \\
        = & \frac{U}{\sqrt{L}} \left[ {_{2}G_{\mathrm{e}}(p_f,k_i;\omega)} + {_{2}G_{\mathrm{e}}(k_f,k_i;\omega)} \right] \nonumber \\
          & \times {_{2}G^0_{\mathrm{w}}}(k_f,p_f;\omega),
\end{align}
along with a straightforward interpretation. A free photon and an excited TLS are initialized 
and propagate until the TLS emits a photon, which leaves two free photons in the waveguide.

\section{Equal-time Green's functions}
\label{sec:appendix:EqualTimeGF}
We will discuss in this Appendix the appearance of equal-time Green's functions in the perturbation 
series for the TLS-Green's function in the two-excitation sector. Our discussion will focus on an 
example diagram of the perturbation series, which is given by

\begin{equation}
 g(t_f-t_i) = \showgraph{height=1cm}{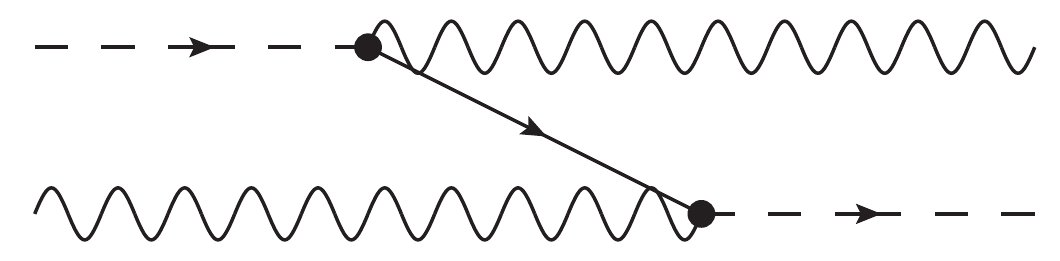},
\end{equation}
with the goal to derive the equal-time Green's functions from a more ``standard'' diagrammatic approach
\cite{Mahan00}. 
Note that this appendix is not mathematically rigorous but intends to give the reader insight about 
the emergence of the equal-time Green's functions.

>From a direct diagrammatic derivation, the diagram is given by

\begin{align}\label{eq:appendix:EqualTimeGF:gInit}
 g(k,k';t_f-&t_i)  \nonumber \\
        = i \frac{U^2}{L} &\int \mathrm{d}t \mathrm{d}t' g_{\mathrm{e}}(t-t_i) g_{\mathrm{ph}}(k,t'-t_i) g_{\mathrm{g}}(t'-t) \nonumber \\
                          &\hphantom{\int \mathrm{d}t \mathrm{d}t'} \times g_{\mathrm{ph}}(k',t_f-t) g_{\mathrm{e}}(t_f-t'),
\end{align}
where the free Green's functions are

\begin{equation}
 g_{\mathrm{e}}(t)=e^{-i\Omega t/2}, \quad g_{\mathrm{g}}(t)=e^{i\Omega t/2}, \quad g_{\mathrm{ph}}(k,t)=e^{-i\epsilon(k)t},
\end{equation}
the according Feynman diagrams are given in Tab.~\ref{tab:FeynmanPresentations} and the time 
integrations stem from the interaction vertices. The crucial point about these Green's functions 
is that they can be rewritten in the form

\begin{align}
 g_{\mathrm{ph}}(k,t_f-t_i) &= e^{-i\epsilon(k)(t_f-t_i)} \nonumber \\
                   &=e^{-i\epsilon(k)(t_f-t)}e^{-i\epsilon(k)(t-t_i)} \nonumber \\
                   &=g_{\mathrm{ph}}(k,t_f-t)g_{\mathrm{ph}}(k,t-t_i).
\end{align}
In this way Eq.~\eqref{eq:appendix:EqualTimeGF:gInit} can be rewritten as

\begin{align}
 g(k,k';t_f-t_i) =&   \nonumber \\
        = i \frac{U^2}{L} \int \mathrm{d}t \mathrm{d}t' &\left[ g_{\mathrm{e}}(t-t_i) g_{\mathrm{ph}}(k,t-t_i) \right] \nonumber \\
                          \times &\left[ g_{\mathrm{g}}(t'-t) g_{\mathrm{ph}}(k,t'-t) g_{\mathrm{ph}}(k',t'-t) \right] \nonumber \\
                          \times &\left[ g_{\mathrm{e}}(t_f-t')g_{\mathrm{ph}}(k',t_f-t') \right],
\end{align}
where each bracket contains an equal-time Green's function. Diagrammatically, this expression can 
be depicted as

\begin{equation}
  g(t_f-t_i) = \showgraph{height=1cm}{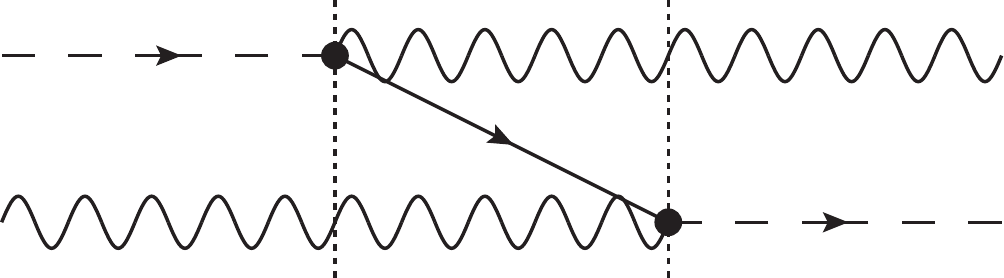},
\end{equation}
which is exactly the form as in Sec.~\ref{sec:FeynmanDiagrams:2pcase}.

\section{Two-excitation scattering matrix}
\label{sec:appendix:2pSMatrix}
In this Appendix, we provide the details of the calculation of the two-excitation scattering 
matrix for the case of a linear dispersion relation. The two-excitation $S$-Matrix is given by 
Eqs. \eqref{eq:2pcase:Linear:SMatrix_def} and \eqref{eq:2pcase:Linear:TMatrix}. With the help 
of Eqs. \eqref{eq:pathint:2pcase:Gd_sym} and \eqref{eq:2pcase:Linear:Gd}, we can write this 
$S$-Matrix as
\begin{equation}\label{eq:appendix:2pSMatrix:Smatrix_decomposition}
 S_{k_{i}p_{i},k_{f}p_{f}} = S^{0}_{k_{i}p_{i},k_{f}p_{f}} + S^{1}_{k_{i}p_{i},k_{f}p_{f}} + S^{2}_{k_{i}p_{i},k_{f}p_{f}},
\end{equation}
\begin{equation}\label{eq:appendix:2pSMatrix:Smatrix_0}
 S^{0}_{k_{i}p_{i},k_{f}p_{f}} = \delta_{k_{i},k_{f}} \delta_{p_{i},p_{f}} + \delta_{p_{i},k_{f}} \delta_{k_{i},p_{f}},
\end{equation}
\begin{align}\label{eq:appendix:2pSMatrix:Smatrix_1}
 S^{1}_{k_{i}p_{i},k_{f}p_{f}} = & -i \delta(\epsilon(k_{i}) + \epsilon(p_{i}) - \epsilon(k_{f}) - \epsilon(p_{f})) \nonumber \\
                                 & \times U^2 \bigl( {_{2}G_{\mathrm{e},\mathrm{r}}} (k_{i};\omega) \delta_{k_{i}, k_{f}} + \{ \text{perm} \} \bigr)\bigr|_{os},
\end{align}
\begin{align}\label{eq:appendix:2pSMatrix:Smatrix_2}
 S^{2}_{k_{i}p_{i},k_{f}p_{f}} = & - i \delta(\epsilon(k_{i}) + \epsilon(p_{i})- \epsilon(k_{f}) - \epsilon(p_{f})) \nonumber \\
                                 & \times \frac{U^4}{2\pi}  \bigl( {_{2}G_{\mathrm{e},\mathrm{r}}} (k_{i}) \, {_{2}G_{\mathrm{w},0}} (k_{i},k_{f}) \, {_{2}G_{\mathrm{e},\mathrm{r}}} (k_{f}) \nonumber \\
                                 & \hphantom{\times \frac{U^4}{2\pi} \bigl(} + \{ \text{perm} \} \bigr)\bigr|_{os},
\end{align}
where $\{ \text{perm} \}$ represents terms where the momenta have been permuted according to 
Eq.~\eqref{eq:pathint:2pcase:Gd_sym}. Furthermore, we suppress the chirality indices, thus 
(initially) treating all (incoming and outgoing) photons as right-movers. The case of different 
chiralities will be discussed at the end of this appendix.

In order to calculate $S^{1}_{k_{i}p_{i},k_{f}p_{f}}$, we rewrite the $\delta$-function 
according to
\begin{equation}\label{eq:appendix:2pSMatrix:deltaEtodeltak}
 \delta(\epsilon(k_{i}) + \epsilon(p_{i}) - \epsilon(k_{f}) - \epsilon(p_{f})) = \frac{\delta_{k_i + p_i, k_f + p_f}}{v}.
\end{equation}
Bearing in mind that we have shifted $\omega \rightarrow \omega - \Omega/2$, using Eq.~\eqref{eq:pathint:2pcase:Gdr},
and setting $\omega= v k_i + v p_i $, we find after cumbersome but straightforward calculation
\begin{align}\label{eq:appendix:2pSMatrix:S1}
 S^{1}_{k_{i}p_{i},k_{f}p_{f}} & = \delta_{p_i, p_f} \delta_{k_i, k_f} \frac{-i U^2 /v}{ v p_i - \Omega +i U^2/v } + \{ \text{perm} \} \nonumber \\
                               & = r_{k_i} \left( \delta_{p_i, p_f} \delta_{k_i, k_f} + \delta_{p_i, k_f} \delta_{k_i, p_f}\right) \nonumber\\
                               & \hphantom{=} + r_{p_i} \left( \delta_{p_i, p_f} \delta_{k_i, k_f} + \delta_{p_i, k_f} \delta_{k_i, p_f}\right).
\end{align}
Here, $r_k$ is the single-excitation reflection amplitude as specified in Eq.~\eqref{eq:spcase:rk_lin}.

In very much the same manner, we find
\begin{align}\label{eq:appendix:2pSMatrix:S2_noDirac}
 S^{2}_{k_{i}p_{i},k_{f}p_{f}} = & \delta_{k_i + p_i, k_f + p_f} 
                                   \frac{-i U^4}{2\pi v} \frac{1}{v k_i - \Omega + i U^2/v} \nonumber \\
                                 & \times \frac{1}{v k_f - v p_i + i 0} 
                                   \frac{1}{v p_f - \Omega + i U^2/v} + \{ \text{perm} \}.
\end{align}

Upon replacing the inner free Green's function by an application of the Dirac identity
\begin{equation}
 \frac{1}{x + i0} = \mathcal{P} \left( \frac{1}{x} \right) - i\pi \delta(x),
\end{equation}
the scattering matrix decomposes into two terms
\begin{equation}
 S^{2}_{k_{i}p_{i},k_{f}p_{f}} = S^{2,\text{P.V.}}_{k_{i}p_{i},k_{f}p_{f}} + S^{2,\delta}_{k_{i}p_{i},k_{f}p_{f}},
\end{equation}
where the term that results from the $\delta$-function in the Dirac identity, $S^{2,\delta}_{k_{i}p_{i},k_{f}p_{f}}$, 
reduces to
\begin{align}\label{eq:appendix:2pSMatrix:S2delta}
 S^{2,\delta}_{k_{i}p_{i},k_{f}p_{f}} = & \frac{1}{2} \delta_{p_i,p_f} \delta_{k_i,k_f} 
                                          \frac{-i U^2/v}{v k_i - \Omega + i U^2 /v} \nonumber \\
                                        & \times \frac{-i U^2/v}{v p_f - \Omega + i U^2 /v} 
                                          + \{ \text{perm} \} \nonumber \\
                                      = & r_{k_1} r_{p_1} \left( \delta_{p_1,p_2} \delta_{k_1,k_2} 
                                          + \delta_{k_1,p_2} \delta_{p_1,k_2} \right).
\end{align}
The term of the scattering matrix that originates from the principal value in the Dirac identity
is given by
\begin{align}
 S^{2,\text{P.V.}}_{k_{i}p_{i},k_{f}p_{f}} = & \frac{i}{2\pi} \delta_{k_i+p_i,k_f+p_f} r_{k_i}r_{p_f} 
                                               \mathcal{P}\frac{1}{k_i - k_f} + \{ \text{perm} \} \nonumber \\
                                           = & \frac{i}{2\pi} \delta_{k_i+p_i,k_f+p_f} 
                                               \biggl[ \mathcal{P}\frac{1}{k_i - k_f} \left( r_{k_i}r_{p_f} 
                                                                                             - r_{p_i}r_{k_f} \right) \nonumber \\
                                             & + \mathcal{P}\frac{1}{k_i - p_f} \left( r_{k_i}r_{k_f} 
                                                                                      - r_{p_i}r_{p_f} \right) 
                                               \biggr].
\end{align}
Here, we have used energy conservation to facilitate certain simplifications. Using energy 
conservation another time, we find
\begin{align}
 r_{k_i}r_{p_f} - & r_{p_i}r_{k_f} \nonumber \\
                = & \frac{U^4}{v} 
                    \frac{(k_i - k_f)\left( E - 2 \Omega + i U^2/v \right)}{(v p_i - \Omega +i U^2/v)(v k_i - \Omega +i U^2/v)} \nonumber \\
                  & \times \frac{1}{(v p_f - \Omega +i U^2/v)(v k_f - \Omega +i U^2/v)}
\end{align}
and the same expression with interchanged momenta, $k_f \leftrightarrow p_f$, for 
$r_{k_i}r_{k_f} - r_{p_i}r_{p_f}$. Exploiting the fact that
\begin{equation}
 \left( k_i - k_f \right) \mathcal{P} \frac{1}{k_i-k_f} = 1
\end{equation}
and combining the above expressions, we find
\begin{align}\label{eq:appendix:2pSMatrix:S2PV}
 S^{2,\text{P.V.}}_{k_{i}p_{i},k_{f}p_{f}} = & \frac{iU^4}{\pi v}\delta_{k_i+p_i,k_f+p_f} \nonumber \\ 
                                             & \times \frac{\left( k_i + p_i - 2 \Omega + i U^2/v \right)}{(v p_i - \Omega +i U^2/v)(v k_i - \Omega +i U^2/v)} \nonumber \\
                                             & \times \frac{1}{(v p_f - \Omega +i U^2/v)(v k_f - \Omega +i U^2/v)}.
\end{align}
Upon inserting Eq.~\eqref{eq:appendix:2pSMatrix:Smatrix_0}, \eqref{eq:appendix:2pSMatrix:S1}, and \eqref{eq:appendix:2pSMatrix:S2delta} 
in \eqref{eq:appendix:2pSMatrix:Smatrix_decomposition}, we finally find
\begin{align}
 S^{RR,RR}_{k_{i}p_{i},k_{f}p_{f}} = & t_{k_i} t_{p_i} \left( \delta_{k_i,k_f} \delta_{p_i,p_f} + 
                                                              \delta_{k_i,p_f} \delta_{p_i,k_f} \right) \nonumber \\
                                     & + S^{2,\text{P.V.}}_{k_{i}p_{i},k_{f}p_{f}},
\end{align}
where $t_k = 1 + r_k$. As a matter of fact, the above expression is exactly the scattering 
matrix given in Ref. \onlinecite{ShenFan-2007-1}.

We now turn to the effects of chirality. First, the momenta are renormalized $k \rightarrow \mu k$, 
which controls the sign of the momenta. Second, chirality is conserved upon free propagation, which 
adds two chirality conserving $\delta$-functions to $S^{0}_{k_{i}p_{i},k_{f}p_{f}}$ and one to 
$S^{1}_{k_{i}p_{i},k_{f}p_{f}}$. Combining all the relevant expressions, the $S$-matrix in the 
other chirality sectors calculates to
\begin{itemize}
 \item $ k_{i}^{R} p_{i}^{R} \rightarrow k_{f}^{R} p_{f}^{L} $
 \begin{align}
  S^{RR,RL}_{k_{i}p_{i},k_{f}p_{f}} = & t_{k_i} r_{p_i} \delta_{k_i,k_f} \delta_{p_i,-p_f} 
                                        + r_{k_i} t_{p_i} \delta_{k_i,-p_f} \delta_{p_i,k_f}  \nonumber \\
                                      & + S^{2,\text{P.V.}}_{k_{i}p_{i},k_{f}p_{f}},
 \end{align}
 \item $ k_{i}^{R} p_{i}^{R} \rightarrow k_{f}^{L} p_{f}^{L} $
 \begin{align}
  S^{RR,LL}_{k_{i}p_{i},k_{f}p_{f}} = & r_{k_i} r_{p_i} \left( \delta_{k_i,-k_f} \delta_{p_i,-p_f} + 
                                                               \delta_{k_i,-p_f} \delta_{p_i,-k_f} \right) \nonumber \\
                                      & + S^{2,\text{P.V.}}_{k_{i}p_{i},k_{f}p_{f}},
 \end{align}
\end{itemize}
where the superscripts of $k_{j}^{\mu}$ indicate the values of chirality.

\bibliography{GreensFunctionCQED}

\begin{thebibliography}{46}
\expandafter\ifx\csname natexlab\endcsname\relax\def\natexlab#1{#1}\fi
\expandafter\ifx\csname bibnamefont\endcsname\relax
  \def\bibnamefont#1{#1}\fi
\expandafter\ifx\csname bibfnamefont\endcsname\relax
  \def\bibfnamefont#1{#1}\fi
\expandafter\ifx\csname citenamefont\endcsname\relax
  \def\citenamefont#1{#1}\fi
\expandafter\ifx\csname url\endcsname\relax
  \def\url#1{\texttt{#1}}\fi
\expandafter\ifx\csname urlprefix\endcsname\relax\def\urlprefix{URL }\fi
\providecommand{\bibinfo}[2]{#2}
\providecommand{\eprint}[2][]{\url{#2}}

\bibitem[{\citenamefont{O'Brien et~al.}(2009)\citenamefont{O'Brien, Furusawa,
  and Vu\v{c}kovi\'{c}}}]{OBrien-2009}
\bibinfo{author}{\bibfnamefont{J.~L.} \bibnamefont{O'Brien}},
  \bibinfo{author}{\bibfnamefont{A.}~\bibnamefont{Furusawa}}, \bibnamefont{and}
  \bibinfo{author}{\bibfnamefont{J.}~\bibnamefont{Vu\v{c}kovi\'{c}}},
  \bibinfo{journal}{Nature Photonics} \textbf{\bibinfo{volume}{3}},
  \bibinfo{pages}{687} (\bibinfo{year}{2009}).

\bibitem[{\citenamefont{Benson}(2011)}]{Benson-2011}
\bibinfo{author}{\bibfnamefont{O.}~\bibnamefont{Benson}},
  \bibinfo{journal}{Nature} \textbf{\bibinfo{volume}{480}},
  \bibinfo{pages}{193} (\bibinfo{year}{2011}).

\bibitem[{\citenamefont{Schell et~al.}(2013)\citenamefont{Schell, Kaschke,
  Fischer, Henze, Wolters, Wegener, and O.}}]{Benson-Wegener-2013}
\bibinfo{author}{\bibfnamefont{A.~W.} \bibnamefont{Schell}},
  \bibinfo{author}{\bibfnamefont{J.}~\bibnamefont{Kaschke}},
  \bibinfo{author}{\bibfnamefont{J.}~\bibnamefont{Fischer}},
  \bibinfo{author}{\bibfnamefont{R.}~\bibnamefont{Henze}},
  \bibinfo{author}{\bibfnamefont{J.}~\bibnamefont{Wolters}},
  \bibinfo{author}{\bibfnamefont{M.}~\bibnamefont{Wegener}}, \bibnamefont{and}
  \bibinfo{author}{\bibfnamefont{B.}~\bibnamefont{O.}},
  \bibinfo{journal}{Scientific Reports} \textbf{\bibinfo{volume}{3}},
  \bibinfo{pages}{1577} (\bibinfo{year}{2013}).

\bibitem[{\citenamefont{You and Nori}(2011)}]{Nori-2011}
\bibinfo{author}{\bibfnamefont{J.~Q.} \bibnamefont{You}} \bibnamefont{and}
  \bibinfo{author}{\bibfnamefont{F.}~\bibnamefont{Nori}},
  \bibinfo{journal}{Nature} \textbf{\bibinfo{volume}{474}},
  \bibinfo{pages}{589} (\bibinfo{year}{2011}).

\bibitem[{\citenamefont{Devoret and Schoelkopf}(2013)}]{Devoret-2013}
\bibinfo{author}{\bibfnamefont{M.~H.} \bibnamefont{Devoret}} \bibnamefont{and}
  \bibinfo{author}{\bibfnamefont{J.~R.} \bibnamefont{Schoelkopf}},
  \bibinfo{journal}{Science} \textbf{\bibinfo{volume}{339}},
  \bibinfo{pages}{1169} (\bibinfo{year}{2013}).

\bibitem[{\citenamefont{Mitsch et~al.}(2014)\citenamefont{Mitsch, Sayrin,
  Albrecht, Schneeweiss, and Rauschenbeutel}}]{Mitsch-2014}
\bibinfo{author}{\bibfnamefont{R.}~\bibnamefont{Mitsch}},
  \bibinfo{author}{\bibfnamefont{C.}~\bibnamefont{Sayrin}},
  \bibinfo{author}{\bibfnamefont{B.}~\bibnamefont{Albrecht}},
  \bibinfo{author}{\bibfnamefont{P.}~\bibnamefont{Schneeweiss}},
  \bibnamefont{and}
  \bibinfo{author}{\bibfnamefont{A.}~\bibnamefont{Rauschenbeutel}},
  \bibinfo{journal}{Nature Communications} \textbf{\bibinfo{volume}{5}},
  \bibinfo{pages}{5713} (\bibinfo{year}{2014}).

\bibitem[{\citenamefont{Shields}(2007)}]{Shields-2007}
\bibinfo{author}{\bibfnamefont{A.~J.} \bibnamefont{Shields}},
  \bibinfo{journal}{Nature Photonics} \textbf{\bibinfo{volume}{1}},
  \bibinfo{pages}{215} (\bibinfo{year}{2007}).

\bibitem[{\citenamefont{Claudon et~al.}(2010)\citenamefont{Claudon, Bleuse,
  Malik, Bazin, Jaffrennou, Gregersen, Sauvan, Lalanne, and
  Gerard}}]{Claudon-2010}
\bibinfo{author}{\bibfnamefont{J.}~\bibnamefont{Claudon}},
  \bibinfo{author}{\bibfnamefont{J.}~\bibnamefont{Bleuse}},
  \bibinfo{author}{\bibfnamefont{N.~S.} \bibnamefont{Malik}},
  \bibinfo{author}{\bibfnamefont{M.}~\bibnamefont{Bazin}},
  \bibinfo{author}{\bibfnamefont{P.}~\bibnamefont{Jaffrennou}},
  \bibinfo{author}{\bibfnamefont{N.}~\bibnamefont{Gregersen}},
  \bibinfo{author}{\bibfnamefont{C.}~\bibnamefont{Sauvan}},
  \bibinfo{author}{\bibfnamefont{P.}~\bibnamefont{Lalanne}}, \bibnamefont{and}
  \bibinfo{author}{\bibfnamefont{J.-M.} \bibnamefont{Gerard}},
  \bibinfo{journal}{Nature Photonics} \textbf{\bibinfo{volume}{4}},
  \bibinfo{pages}{174} (\bibinfo{year}{2010}).

\bibitem[{\citenamefont{Laucht et~al.}(2012)\citenamefont{Laucht, P{\"u}tz,
  G{\"u}nthner, Hauke, Saive, Fr\'{e}d\'{e}rick, Bichler, Amann, Holleitner,
  Kaniber et~al.}}]{Laucht-2012}
\bibinfo{author}{\bibfnamefont{A.}~\bibnamefont{Laucht}},
  \bibinfo{author}{\bibfnamefont{S.}~\bibnamefont{P{\"u}tz}},
  \bibinfo{author}{\bibfnamefont{T.}~\bibnamefont{G{\"u}nthner}},
  \bibinfo{author}{\bibfnamefont{N.}~\bibnamefont{Hauke}},
  \bibinfo{author}{\bibfnamefont{R.}~\bibnamefont{Saive}},
  \bibinfo{author}{\bibfnamefont{S.}~\bibnamefont{Fr\'{e}d\'{e}rick}},
  \bibinfo{author}{\bibfnamefont{M.}~\bibnamefont{Bichler}},
  \bibinfo{author}{\bibfnamefont{M.-C.} \bibnamefont{Amann}},
  \bibinfo{author}{\bibfnamefont{A.~W.} \bibnamefont{Holleitner}},
  \bibinfo{author}{\bibfnamefont{M.}~\bibnamefont{Kaniber}},
  \bibnamefont{et~al.}, \bibinfo{journal}{Phys. Rev. X}
  \textbf{\bibinfo{volume}{2}}, \bibinfo{pages}{011014} (\bibinfo{year}{2012}).

\bibitem[{\citenamefont{Fattahpoor et~al.}(2013)\citenamefont{Fattahpoor,
  Hoang, Midolo, Dietrich, Li, Linfield, Schouwenberg, Xia, Pagliano, van Otten
  et~al.}}]{Fattahpoor-2013}
\bibinfo{author}{\bibfnamefont{S.}~\bibnamefont{Fattahpoor}},
  \bibinfo{author}{\bibfnamefont{T.~B.} \bibnamefont{Hoang}},
  \bibinfo{author}{\bibfnamefont{L.}~\bibnamefont{Midolo}},
  \bibinfo{author}{\bibfnamefont{C.~P.} \bibnamefont{Dietrich}},
  \bibinfo{author}{\bibfnamefont{L.~H.} \bibnamefont{Li}},
  \bibinfo{author}{\bibfnamefont{E.~H.} \bibnamefont{Linfield}},
  \bibinfo{author}{\bibfnamefont{J.~F.~P.} \bibnamefont{Schouwenberg}},
  \bibinfo{author}{\bibfnamefont{T.}~\bibnamefont{Xia}},
  \bibinfo{author}{\bibfnamefont{F.~M.} \bibnamefont{Pagliano}},
  \bibinfo{author}{\bibfnamefont{F.~M.~W.} \bibnamefont{van Otten}},
  \bibnamefont{et~al.}, \bibinfo{journal}{Appl. Phys. Lett.}
  \textbf{\bibinfo{volume}{102}}, \bibinfo{pages}{131105}
  (\bibinfo{year}{2013}).

\bibitem[{\citenamefont{Hadfield}(2009)}]{Hadfield-2009}
\bibinfo{author}{\bibfnamefont{R.~H.} \bibnamefont{Hadfield}},
  \bibinfo{journal}{Nature Photonics} \textbf{\bibinfo{volume}{3}},
  \bibinfo{pages}{696} (\bibinfo{year}{2009}).

\bibitem[{\citenamefont{Pernice et~al.}(2012)\citenamefont{Pernice, Schuck,
  Minaeva, Li, Goltsman, Sergienko, and Tang}}]{Pernice-2012}
\bibinfo{author}{\bibfnamefont{W.~H.~P.} \bibnamefont{Pernice}},
  \bibinfo{author}{\bibfnamefont{C.}~\bibnamefont{Schuck}},
  \bibinfo{author}{\bibfnamefont{O.}~\bibnamefont{Minaeva}},
  \bibinfo{author}{\bibfnamefont{M.}~\bibnamefont{Li}},
  \bibinfo{author}{\bibfnamefont{G.~N.} \bibnamefont{Goltsman}},
  \bibinfo{author}{\bibfnamefont{A.~V.} \bibnamefont{Sergienko}},
  \bibnamefont{and} \bibinfo{author}{\bibfnamefont{H.~X.} \bibnamefont{Tang}},
  \bibinfo{journal}{Nature Communications} \textbf{\bibinfo{volume}{3}},
  \bibinfo{pages}{1325} (\bibinfo{year}{2012}).

\bibitem[{\citenamefont{Sahin et~al.}(2013)\citenamefont{Sahin, Gaggero, Zhou,
  Jahanmirinejad, Mattioli, Leoni, Beetz, Lermer, Kamp, H{\"o}fling
  et~al.}}]{Sahin-2013}
\bibinfo{author}{\bibfnamefont{D.}~\bibnamefont{Sahin}},
  \bibinfo{author}{\bibfnamefont{A.}~\bibnamefont{Gaggero}},
  \bibinfo{author}{\bibfnamefont{Z.}~\bibnamefont{Zhou}},
  \bibinfo{author}{\bibfnamefont{S.}~\bibnamefont{Jahanmirinejad}},
  \bibinfo{author}{\bibfnamefont{F.}~\bibnamefont{Mattioli}},
  \bibinfo{author}{\bibfnamefont{R.}~\bibnamefont{Leoni}},
  \bibinfo{author}{\bibfnamefont{J.}~\bibnamefont{Beetz}},
  \bibinfo{author}{\bibfnamefont{M.}~\bibnamefont{Lermer}},
  \bibinfo{author}{\bibfnamefont{M.}~\bibnamefont{Kamp}},
  \bibinfo{author}{\bibfnamefont{S.}~\bibnamefont{H{\"o}fling}},
  \bibnamefont{et~al.}, \bibinfo{journal}{Appl. Phys. Lett.}
  \textbf{\bibinfo{volume}{103}}, \bibinfo{pages}{111116}
  (\bibinfo{year}{2013}).

\bibitem[{\citenamefont{Najafi et~al.}(2015)\citenamefont{Najafi, Mowver,
  Harris, Bellei, Dane, Lee, Hu, Kharel, Marsili, Assefa et~al.}}]{Najafi-2015}
\bibinfo{author}{\bibfnamefont{F.}~\bibnamefont{Najafi}},
  \bibinfo{author}{\bibfnamefont{J.}~\bibnamefont{Mowver}},
  \bibinfo{author}{\bibfnamefont{N.~C.} \bibnamefont{Harris}},
  \bibinfo{author}{\bibfnamefont{F.}~\bibnamefont{Bellei}},
  \bibinfo{author}{\bibfnamefont{A.}~\bibnamefont{Dane}},
  \bibinfo{author}{\bibfnamefont{C.}~\bibnamefont{Lee}},
  \bibinfo{author}{\bibfnamefont{X.}~\bibnamefont{Hu}},
  \bibinfo{author}{\bibfnamefont{P.}~\bibnamefont{Kharel}},
  \bibinfo{author}{\bibfnamefont{F.}~\bibnamefont{Marsili}},
  \bibinfo{author}{\bibfnamefont{S.}~\bibnamefont{Assefa}},
  \bibnamefont{et~al.}, \bibinfo{journal}{Nature Communications}
  \textbf{\bibinfo{volume}{6}}, \bibinfo{pages}{5873} (\bibinfo{year}{2015}).

\bibitem[{\citenamefont{Chang et~al.}(2014)\citenamefont{Chang, Vuleti\'{c},
  and Lukin}}]{Chang-2014}
\bibinfo{author}{\bibfnamefont{D.~E.} \bibnamefont{Chang}},
  \bibinfo{author}{\bibfnamefont{V.}~\bibnamefont{Vuleti\'{c}}},
  \bibnamefont{and} \bibinfo{author}{\bibfnamefont{M.~D.} \bibnamefont{Lukin}},
  \bibinfo{journal}{Nature Photonics} \textbf{\bibinfo{volume}{8}},
  \bibinfo{pages}{685} (\bibinfo{year}{2014}).

\bibitem[{\citenamefont{Volz et~al.}(2014)\citenamefont{Volz, Scheucher, Junge,
  and Rauschenbeutel}}]{Volz-2014}
\bibinfo{author}{\bibfnamefont{J.}~\bibnamefont{Volz}},
  \bibinfo{author}{\bibfnamefont{M.}~\bibnamefont{Scheucher}},
  \bibinfo{author}{\bibfnamefont{C.}~\bibnamefont{Junge}}, \bibnamefont{and}
  \bibinfo{author}{\bibfnamefont{A.}~\bibnamefont{Rauschenbeutel}},
  \bibinfo{journal}{Nature Photonics} \textbf{\bibinfo{volume}{8}},
  \bibinfo{pages}{965} (\bibinfo{year}{2014}).

\bibitem[{\citenamefont{Lodahl et~al.}(2015)\citenamefont{Lodahl, Mahmoodian,
  and Stobbe}}]{Lodahl-2015}
\bibinfo{author}{\bibfnamefont{P.}~\bibnamefont{Lodahl}},
  \bibinfo{author}{\bibfnamefont{S.}~\bibnamefont{Mahmoodian}},
  \bibnamefont{and} \bibinfo{author}{\bibfnamefont{S.}~\bibnamefont{Stobbe}},
  \bibinfo{journal}{Rev. Mod. Phys.} \textbf{\bibinfo{volume}{87}},
  \bibinfo{pages}{347} (\bibinfo{year}{2015}).

\bibitem[{\citenamefont{Shen and Fan}(2007{\natexlab{a}})}]{ShenFan-2007-1}
\bibinfo{author}{\bibfnamefont{J.-T.} \bibnamefont{Shen}} \bibnamefont{and}
  \bibinfo{author}{\bibfnamefont{S.}~\bibnamefont{Fan}},
  \bibinfo{journal}{Phys. Rev. A} \textbf{\bibinfo{volume}{76}},
  \bibinfo{pages}{062709} (\bibinfo{year}{2007}{\natexlab{a}}).

\bibitem[{\citenamefont{Shi and Sun}(2009)}]{ShiSun-2009}
\bibinfo{author}{\bibfnamefont{T.}~\bibnamefont{Shi}} \bibnamefont{and}
  \bibinfo{author}{\bibfnamefont{C.~P.} \bibnamefont{Sun}},
  \bibinfo{journal}{Phys. Rev. B} \textbf{\bibinfo{volume}{79}},
  \bibinfo{pages}{205111} (\bibinfo{year}{2009}).

\bibitem[{\citenamefont{Fan et~al.}(2010)\citenamefont{Fan, Kocaba\c{s}, and
  Shen}}]{Fan10}
\bibinfo{author}{\bibfnamefont{S.}~\bibnamefont{Fan}},
  \bibinfo{author}{\bibfnamefont{S.~E.} \bibnamefont{Kocaba\c{s}}},
  \bibnamefont{and} \bibinfo{author}{\bibfnamefont{J.-T.} \bibnamefont{Shen}},
  \bibinfo{journal}{Phys. Rev. A} \textbf{\bibinfo{volume}{82}},
  \bibinfo{pages}{063821} (\bibinfo{year}{2010}).

\bibitem[{\citenamefont{Pletyukhov and Gritsev}(2012)}]{Pletyukhov-2012}
\bibinfo{author}{\bibfnamefont{M.}~\bibnamefont{Pletyukhov}} \bibnamefont{and}
  \bibinfo{author}{\bibfnamefont{V.}~\bibnamefont{Gritsev}},
  \bibinfo{journal}{New Journal of Physics} \textbf{\bibinfo{volume}{14}},
  \bibinfo{pages}{095028} (\bibinfo{year}{2012}).

\bibitem[{\citenamefont{Zheng et~al.}(2010)\citenamefont{Zheng, Gauthier, and
  Baranger}}]{Zheng10}
\bibinfo{author}{\bibfnamefont{H.}~\bibnamefont{Zheng}},
  \bibinfo{author}{\bibfnamefont{D.~J.} \bibnamefont{Gauthier}},
  \bibnamefont{and} \bibinfo{author}{\bibfnamefont{H.~U.}
  \bibnamefont{Baranger}}, \bibinfo{journal}{Phys. Rev. A}
  \textbf{\bibinfo{volume}{82}}, \bibinfo{pages}{063816}
  (\bibinfo{year}{2010}).

\bibitem[{\citenamefont{Zheng et~al.}(2011)\citenamefont{Zheng, Gauthier, and
  Baranger}}]{Zheng11}
\bibinfo{author}{\bibfnamefont{H.}~\bibnamefont{Zheng}},
  \bibinfo{author}{\bibfnamefont{D.~J.} \bibnamefont{Gauthier}},
  \bibnamefont{and} \bibinfo{author}{\bibfnamefont{H.~U.}
  \bibnamefont{Baranger}}, \bibinfo{journal}{Phys. Rev. Lett.}
  \textbf{\bibinfo{volume}{107}}, \bibinfo{pages}{223601}
  (\bibinfo{year}{2011}).

\bibitem[{\citenamefont{Zheng et~al.}(2012)\citenamefont{Zheng, Gauthier, and
  Baranger}}]{Zheng12}
\bibinfo{author}{\bibfnamefont{H.}~\bibnamefont{Zheng}},
  \bibinfo{author}{\bibfnamefont{D.~J.} \bibnamefont{Gauthier}},
  \bibnamefont{and} \bibinfo{author}{\bibfnamefont{H.~U.}
  \bibnamefont{Baranger}}, \bibinfo{journal}{Phys. Rev. A}
  \textbf{\bibinfo{volume}{85}}, \bibinfo{pages}{043832}
  (\bibinfo{year}{2012}).

\bibitem[{\citenamefont{Longo et~al.}(2009)\citenamefont{Longo, Schmitteckert,
  and Busch}}]{Longo-2009}
\bibinfo{author}{\bibfnamefont{P.}~\bibnamefont{Longo}},
  \bibinfo{author}{\bibfnamefont{P.}~\bibnamefont{Schmitteckert}},
  \bibnamefont{and} \bibinfo{author}{\bibfnamefont{K.}~\bibnamefont{Busch}},
  \bibinfo{journal}{J. Opt. A} \textbf{\bibinfo{volume}{11}},
  \bibinfo{pages}{114009} (\bibinfo{year}{2009}).

\bibitem[{\citenamefont{Longo et~al.}(2010)\citenamefont{Longo, Schmitteckert,
  and Busch}}]{LongoSchmitteckertBusch-2010}
\bibinfo{author}{\bibfnamefont{P.}~\bibnamefont{Longo}},
  \bibinfo{author}{\bibfnamefont{P.}~\bibnamefont{Schmitteckert}},
  \bibnamefont{and} \bibinfo{author}{\bibfnamefont{K.}~\bibnamefont{Busch}},
  \bibinfo{journal}{Phys. Rev. Lett.} \textbf{\bibinfo{volume}{104}},
  \bibinfo{pages}{023602} (\bibinfo{year}{2010}).

\bibitem[{\citenamefont{Nysteen et~al.}(2015)\citenamefont{Nysteen, McCutcheon,
  and M{\o}rk}}]{Nysteen-2015}
\bibinfo{author}{\bibfnamefont{A.}~\bibnamefont{Nysteen}},
  \bibinfo{author}{\bibfnamefont{D.~P.~S.} \bibnamefont{McCutcheon}},
  \bibnamefont{and} \bibinfo{author}{\bibfnamefont{J.}~\bibnamefont{M{\o}rk}},
  \bibinfo{journal}{New J. Phys.} \textbf{\bibinfo{volume}{17}},
  \bibinfo{pages}{023030} (\bibinfo{year}{2015}).

\bibitem[{\citenamefont{Sanchez-Burillo
  et~al.}(2014)\citenamefont{Sanchez-Burillo, Zueco, Garcia-Ripoll, and
  Martin-Moreno}}]{Burillo-2014}
\bibinfo{author}{\bibfnamefont{E.}~\bibnamefont{Sanchez-Burillo}},
  \bibinfo{author}{\bibfnamefont{D.}~\bibnamefont{Zueco}},
  \bibinfo{author}{\bibfnamefont{J.~J.} \bibnamefont{Garcia-Ripoll}},
  \bibnamefont{and}
  \bibinfo{author}{\bibfnamefont{L.}~\bibnamefont{Martin-Moreno}},
  \bibinfo{journal}{Phys. Rev. Lett.} \textbf{\bibinfo{volume}{113}},
  \bibinfo{pages}{263604} (\bibinfo{year}{2014}).

\bibitem[{\citenamefont{Notomi et~al.}(2008)\citenamefont{Notomi, Kuramochi,
  and Tanabe}}]{Notomi-2008}
\bibinfo{author}{\bibfnamefont{M.}~\bibnamefont{Notomi}},
  \bibinfo{author}{\bibfnamefont{E.}~\bibnamefont{Kuramochi}},
  \bibnamefont{and} \bibinfo{author}{\bibfnamefont{T.}~\bibnamefont{Tanabe}},
  \bibinfo{journal}{Nat. Photon.} \textbf{\bibinfo{volume}{2}},
  \bibinfo{pages}{741} (\bibinfo{year}{2008}).

\bibitem[{\citenamefont{Takesue et~al.}(2013)\citenamefont{Takesue, Matsuda,
  Kuramochi, Munro, and Notomi}}]{Notomi-2013}
\bibinfo{author}{\bibfnamefont{H.}~\bibnamefont{Takesue}},
  \bibinfo{author}{\bibfnamefont{N.}~\bibnamefont{Matsuda}},
  \bibinfo{author}{\bibfnamefont{E.}~\bibnamefont{Kuramochi}},
  \bibinfo{author}{\bibfnamefont{W.~J.} \bibnamefont{Munro}}, \bibnamefont{and}
  \bibinfo{author}{\bibfnamefont{M.}~\bibnamefont{Notomi}},
  \bibinfo{journal}{Nat. Commun.} \textbf{\bibinfo{volume}{4}},
  \bibinfo{eid}{2725} (\bibinfo{year}{2013}).

\bibitem[{\citenamefont{Schell et~al.}(2015)\citenamefont{Schell, Takashima,
  Kamioka, Oe, Fujiwara, O., and Takeuchi}}]{Schell-2015}
\bibinfo{author}{\bibfnamefont{A.~W.} \bibnamefont{Schell}},
  \bibinfo{author}{\bibfnamefont{H.}~\bibnamefont{Takashima}},
  \bibinfo{author}{\bibfnamefont{S.}~\bibnamefont{Kamioka}},
  \bibinfo{author}{\bibfnamefont{Y.}~\bibnamefont{Oe}},
  \bibinfo{author}{\bibfnamefont{M.}~\bibnamefont{Fujiwara}},
  \bibinfo{author}{\bibfnamefont{B.}~\bibnamefont{O.}}, \bibnamefont{and}
  \bibinfo{author}{\bibfnamefont{S.}~\bibnamefont{Takeuchi}},
  \bibinfo{journal}{Scientific Reports} \textbf{\bibinfo{volume}{5}},
  \bibinfo{pages}{9619} (\bibinfo{year}{2015}).

\bibitem[{\citenamefont{Jaynes and Cummings}(1963)}]{JaynesCummings-1963}
\bibinfo{author}{\bibfnamefont{E.}~\bibnamefont{Jaynes}} \bibnamefont{and}
  \bibinfo{author}{\bibfnamefont{F.~W.} \bibnamefont{Cummings}},
  \bibinfo{journal}{Proceedings of the IEEE} \textbf{\bibinfo{volume}{51}},
  \bibinfo{pages}{89} (\bibinfo{year}{1963}), ISSN \bibinfo{issn}{0018-9219}.

\bibitem[{\citenamefont{Dicke}(1954)}]{Dicke-1954}
\bibinfo{author}{\bibfnamefont{R.}~\bibnamefont{Dicke}},
  \bibinfo{journal}{Phys. Rev.} \textbf{\bibinfo{volume}{93}},
  \bibinfo{pages}{99} (\bibinfo{year}{1954}).

\bibitem[{\citenamefont{Anderson}(1961)}]{Anderson-1961}
\bibinfo{author}{\bibfnamefont{P.~W.} \bibnamefont{Anderson}},
  \bibinfo{journal}{Phys. Rev.} \textbf{\bibinfo{volume}{124}},
  \bibinfo{pages}{41} (\bibinfo{year}{1961}).

\bibitem[{\citenamefont{Rammer}(2004)}]{Rammer04}
\bibinfo{author}{\bibfnamefont{J.}~\bibnamefont{Rammer}},
  \emph{\bibinfo{title}{Quantum Transport Theory}}, Frontiers in Physics Series
  (\bibinfo{publisher}{Westview Press}, \bibinfo{year}{2004}), ISBN
  \bibinfo{isbn}{9780813346229}.

\bibitem[{\citenamefont{Rammer}(2007)}]{Rammer07}
\bibinfo{author}{\bibfnamefont{J.}~\bibnamefont{Rammer}},
  \emph{\bibinfo{title}{Quantum Field Theory of Non-equilibrium States}}
  (\bibinfo{publisher}{Cambridge University Press}, \bibinfo{year}{2007}), ISBN
  \bibinfo{isbn}{9781139465014}.

\bibitem[{\citenamefont{Boudjedaa et~al.}(1996)\citenamefont{Boudjedaa,
  Bounames, Nouicer, Chetouani, and
  Hammann}}]{BoudjedaaBounamesNouicerChetouaniHammann_1996}
\bibinfo{author}{\bibfnamefont{T.}~\bibnamefont{Boudjedaa}},
  \bibinfo{author}{\bibfnamefont{A.}~\bibnamefont{Bounames}},
  \bibinfo{author}{\bibfnamefont{K.}~\bibnamefont{Nouicer}},
  \bibinfo{author}{\bibfnamefont{L.}~\bibnamefont{Chetouani}},
  \bibnamefont{and} \bibinfo{author}{\bibfnamefont{T.~F.}
  \bibnamefont{Hammann}}, \bibinfo{journal}{Physica Scripta}
  \textbf{\bibinfo{volume}{54}}, \bibinfo{pages}{225} (\bibinfo{year}{1996}).

\bibitem[{\citenamefont{Moeferdt et~al.}(2013)\citenamefont{Moeferdt,
  Schmitteckert, and Busch}}]{Moeferdt-2013}
\bibinfo{author}{\bibfnamefont{M.}~\bibnamefont{Moeferdt}},
  \bibinfo{author}{\bibfnamefont{P.}~\bibnamefont{Schmitteckert}},
  \bibnamefont{and} \bibinfo{author}{\bibfnamefont{K.}~\bibnamefont{Busch}},
  \bibinfo{journal}{Opt. Lett.} \textbf{\bibinfo{volume}{38}},
  \bibinfo{pages}{3693} (\bibinfo{year}{2013}).

\bibitem[{\citenamefont{Longo et~al.}(2011)\citenamefont{Longo, Schmitteckert,
  and Busch}}]{LongoSchmitteckertBusch-2011}
\bibinfo{author}{\bibfnamefont{P.}~\bibnamefont{Longo}},
  \bibinfo{author}{\bibfnamefont{P.}~\bibnamefont{Schmitteckert}},
  \bibnamefont{and} \bibinfo{author}{\bibfnamefont{K.}~\bibnamefont{Busch}},
  \bibinfo{journal}{Phys. Rev. A} \textbf{\bibinfo{volume}{83}},
  \bibinfo{pages}{063828} (\bibinfo{year}{2011}).

\bibitem[{\citenamefont{Lehmann et~al.}(1955)\citenamefont{Lehmann, Symanzik,
  and Zimmermann}}]{LehmannSymanzikZimmermann-1955}
\bibinfo{author}{\bibfnamefont{H.}~\bibnamefont{Lehmann}},
  \bibinfo{author}{\bibfnamefont{K.}~\bibnamefont{Symanzik}}, \bibnamefont{and}
  \bibinfo{author}{\bibfnamefont{W.}~\bibnamefont{Zimmermann}},
  \bibinfo{journal}{Il Nuovo Cimento} \textbf{\bibinfo{volume}{1}},
  \bibinfo{pages}{205} (\bibinfo{year}{1955}), ISSN \bibinfo{issn}{0029-6341}.

\bibitem[{\citenamefont{Zar\'and et~al.}(2004)\citenamefont{Zar\'and, Borda,
  von Delft, and Andrei}}]{ZarandBordaVonDelftAndrei-2004}
\bibinfo{author}{\bibfnamefont{G.}~\bibnamefont{Zar\'and}},
  \bibinfo{author}{\bibfnamefont{L.}~\bibnamefont{Borda}},
  \bibinfo{author}{\bibfnamefont{J.}~\bibnamefont{von Delft}},
  \bibnamefont{and} \bibinfo{author}{\bibfnamefont{N.}~\bibnamefont{Andrei}},
  \bibinfo{journal}{Phys. Rev. Lett.} \textbf{\bibinfo{volume}{93}},
  \bibinfo{pages}{107204} (\bibinfo{year}{2004}).

\bibitem[{\citenamefont{Borda et~al.}(2007)\citenamefont{Borda, Fritz, Andrei,
  and Zar\'and}}]{BordaFritzAndreiZarand_2007}
\bibinfo{author}{\bibfnamefont{L.}~\bibnamefont{Borda}},
  \bibinfo{author}{\bibfnamefont{L.}~\bibnamefont{Fritz}},
  \bibinfo{author}{\bibfnamefont{N.}~\bibnamefont{Andrei}}, \bibnamefont{and}
  \bibinfo{author}{\bibfnamefont{G.}~\bibnamefont{Zar\'and}},
  \bibinfo{journal}{Phys. Rev. B} \textbf{\bibinfo{volume}{75}},
  \bibinfo{pages}{235112} (\bibinfo{year}{2007}).

\bibitem[{\citenamefont{Zhou et~al.}(2008)\citenamefont{Zhou, Gong, Liu, Sun,
  and Nori}}]{Nori-2008}
\bibinfo{author}{\bibfnamefont{L.}~\bibnamefont{Zhou}},
  \bibinfo{author}{\bibfnamefont{Z.~R.} \bibnamefont{Gong}},
  \bibinfo{author}{\bibfnamefont{Y.-x.} \bibnamefont{Liu}},
  \bibinfo{author}{\bibfnamefont{C.~P.} \bibnamefont{Sun}}, \bibnamefont{and}
  \bibinfo{author}{\bibfnamefont{F.}~\bibnamefont{Nori}},
  \bibinfo{journal}{Phys. Rev. Lett.} \textbf{\bibinfo{volume}{101}},
  \bibinfo{pages}{100501} (\bibinfo{year}{2008}).

\bibitem[{\citenamefont{Braun and
  Schmitteckert}(2014)}]{Braun_Schmitteckert:PRB14}
\bibinfo{author}{\bibfnamefont{A.}~\bibnamefont{Braun}} \bibnamefont{and}
  \bibinfo{author}{\bibfnamefont{P.}~\bibnamefont{Schmitteckert}},
  \bibinfo{journal}{Phys. Rev. B} \textbf{\bibinfo{volume}{90}},
  \bibinfo{pages}{165112} (\bibinfo{year}{2014}).

\bibitem[{\citenamefont{Shen and Fan}(2007{\natexlab{b}})}]{ShenFan-2007}
\bibinfo{author}{\bibfnamefont{J.-T.} \bibnamefont{Shen}} \bibnamefont{and}
  \bibinfo{author}{\bibfnamefont{S.}~\bibnamefont{Fan}},
  \bibinfo{journal}{Phys. Rev. Lett.} \textbf{\bibinfo{volume}{98}},
  \bibinfo{pages}{153003} (\bibinfo{year}{2007}{\natexlab{b}}).

\bibitem[{\citenamefont{Mahan}(2000)}]{Mahan00}
\bibinfo{author}{\bibfnamefont{G.}~\bibnamefont{Mahan}},
  \emph{\bibinfo{title}{Many-Particle Physics}}, Physics of Solids and Liquids
  (\bibinfo{publisher}{Springer}, \bibinfo{year}{2000}), ISBN
  \bibinfo{isbn}{9780306463389}.

\end{thebibliography}

\end{document}